\documentclass[10pt]{iopart}

\makeatletter
\expandafter\let\csname equation*\endcsname\relax
\expandafter\let\csname endequation*\endcsname\relax
\makeatother

\usepackage{amsmath} 

\usepackage{xcolor}
\usepackage{soul} 
\usepackage{graphicx}
\usepackage{mathrsfs}  
\usepackage{amsmath}
\usepackage{float}
\usepackage{url}
\usepackage[numbers,square,sort&compress]{natbib}

\begin{document}

\title{Detachment dynamics and disturbance rejection in
the TCV X-Point Target divertor}

\author{M Winkel$^{1,2}$, K Verhaegh$^1$, B Kool$^{2,3}$, K Lee$^4$,  M Carpita$^4$, D Brida$^5$, G L Derks$^{2}$, O Février$^4$, R I Morgan$^4$, O. Pan$^5$, A Perek$^4$, C Theiler$^4$, M. Zurita$^4$, M van Berkel$^2$, the TCV team$^\dagger$ and the EUROfusion tokamak exploitation team$^\S$}

\makeatletter

\makeatother

\address{$^{1}$Dept. of Applied Physics and Science Communication, Eindhoven University of Technology, Eindhoven, Netherlands}
\address{$^2$DIFFER - Dutch Institute for Fundamental Energy Research, Eindhoven, Netherlands}
\address{$^{3}$Dept. of Mechanical Engineering, Control Systems Technology Group, Eindhoven University of Technology, Eindhoven, Netherlands}
\address{$^4$École Polytechnique Fédérale de Lausanne (EPFL), Swiss Plasma Center (SPC), Lausanne, Switzerland}
\address{$^5$Max-Planck-Institut für Plasmaphysik, Garching bei München, Germany}
\address{$\dagger$ See author list of C. Theiler, et al., Nucl. Fusion 66 (2026) 116007.}
\address{$\S$ See author list of N. Vianello, et al., Nucl. Fusion 66 (2026) 116010.}

\ead{m.w.g.winkel@differ.nl}

\vspace{10pt}

\begin{indented}
\item[]October 2026 
\end{indented}

\begin{abstract}
The X-Point Target (XPT) is an alternative divertor configuration with a secondary X-point in the divertor leg. We investigate its detachment dynamics and disturbance rejection capacity on the TCV tokamak, comparing it with a long-legged single-null (SN) divertor. We employ a system identification approach using multi-sine perturbations to D$_2$ fuelling in Ohmic L-mode scenarios and to N$_2$ seeding and electron cyclotron resonance heating (ECRH) power in auxiliary-heated L-mode scenarios, using the CIII emission front position as a proxy for the detached state evolution. For D$_2$ fuelling perturbations at 4--31~Hz, the gain of the front response at the secondary X-point is at least a factor of 3 lower than upstream of it, where the response is similar to that of the SN. This indicates that the reduced detachment sensitivity of the secondary X-point in an XPT configuration previously observed in quasi-steady state experiments, extends to control-relevant timescales as an enhanced disturbance rejection capacity. In auxiliary-heated scenarios, the CIII front in the XPT remains nearly stationary near the secondary X-point under N$_2$ seeding and ECRH power perturbations, whereas the CIII front in SN discharges responds strongly. However, due to complex SOL profiles in the region close to the secondary X-point of the XPT configuration, this reduction in CIII front response may not imply disturbance rejection of the detached state as a whole. The enhanced disturbance rejection at the secondary X-point could provide a passive buffer against disturbances, simplifying the task of exhaust controllers to maintain the detached state throughout a discharge cycle.
\end{abstract}

\vspace{1pc}

\vspace{1pc}
\noindent{To be submitted to}: \textit{Nuclear Fusion}

%
%
\maketitle
%
\ioptwocol

\section{Introduction}\label{sect: Introduction}
Power exhaust remains one of the key challenges on the path to reactor-grade tokamaks. Without mitigation, the projected target heat loads and plasma temperatures in reactor-class devices like EU-DEMO \cite{Zohm2021DEMOExhaustProblem} and especially compact devices like STEP \cite{Henderson2025StepOverviewPowerExhaust}, SPARC \cite{Creely2020SPARCOverview}, and ARC \cite{Sorbom2015ARCDesign,Eich2026_ARC} can largely exceed the material limits of the divertor targets, posing a risk to structural machine damage. \par
To stay within the wall material limits, detached operation of the divertor plasma is required. In the detached state, the plasma flowing from the core into the scrape-off layer (SOL) first cools through radiative interactions with impurity species, often actively injected, promoting volumetric power dissipation \cite{Gao2023Seeding, Reinke2017Seeding}. Secondly, plasma-neutral interactions further remove power, momentum and particles (ions) from the plasma \cite{Lipschultz1998VolumeRecomb,Krashenninikov2017Detachment,Verhaegh2019PlasmaAtom,Verhaegh2021PlasmaMolecule}, which can be enhanced via the injection of additional hydrogenic species in the divertor chamber. This can cause the ionisation source (\textit{detachment front}) to 'detach' from the target as a neutral particle buffer builds in front of the target surface in the divertor. Consequently, the heat and particle fluxes to the target are significantly reduced. To ensure divertor wall protection, the front must therefore remain detached from the target, yet not approach the core close enough to reduce fusion performance \cite{Lipschultz2007DivertorPhysics, Ravensbergen2021NaturePaper,Theiler2017_FirstADCs} and potentially violent disruptions \cite{ITERbasis1999Disruptions,Hender2007Disruptions,Boozer2012Disruptions}. 
Maintaining the detached state within these two hard limits throughout a discharge cycle is rendered challenging by transient events occurring across multiple timescales, such as core plasma instabilities \cite{Wesson2004Tokamaks}, low (L) to high (H) confinement mode transitions \cite{Wagner1982Hmode} (or vice versa), core pellet fuelling \cite{Wiesen2017Pellets} or actuator failures \cite{Kool2026_STEP}. The situation is further complicated by the limited actuation possibilities for fast disturbances due to the slow actuation of gas valves, as well as significant injection delays caused by long transportation times via gas pipes \cite{VanBerkel2025sysIDOverview}.  \par
With regard to the described exhaust challenges, recent studies of Alternative Divertor Configurations (ADCs) in TCV and MAST-U show key exhaust benefits compared to  conventional divertor geometries in quasi-steady state. Experiments on the Super-X Divertor (SXD) at MAST-U demonstrate reduced target heat fluxes, improved detachment access (i.e. reduced \textit{detachment onset}), an enlarged operational parameter space for which the detachment front is located in between the target and X-point (\textit{detachment window}), and a reduction of the \textit{detachment sensitivity} to core conditions  \cite{Verhaegh2025DivertorShapingPowerExhaust}. Although the geometry of the TCV X-Point Target (XPT) divertor, which includes a secondary X-Point, differs substantially from the SXD, it features similar benefits. Reduced target heat loads, an enlarged detachment window, and indications of reduced detachment front sensitivity near the secondary X point were all observed \cite{Lee2025XPT}.\par
While these recent ADC results show promise for mitigating stationary power exhaust, reactor-relevant solutions must also address \textit{dynamic} behaviour of the detached state throughout a discharge, requiring an understanding of how it responds to disturbances, arising from transients with different timescales. \par
To study this behaviour for control-relevant timescales spanning the ms to s range, a system identification method can be used to identify detachment dynamics, which has been demonstrated in multiple tokamaks \cite{koenders_systematic_2022,Bosman2025AUGandJETSysID,Kool2025NaturePaper,Kool2025DoubleNull,Ceelen2025SysIDonDIIID}.
Using system identification, experiments with sinusoidal gas puff modulations on the SXD at MAST-U confirm that the observed detachment sensitivity reduction in quasi-steady-state extrapolate towards these timescales \cite{Kool2025NaturePaper}. This built-in buffer provides actuators with more time to respond to disturbances and is thus favourable for maintaining the detached state throughout a discharge.\par 
Building on the TCV X-Point Target and MAST-U results, this work addresses the disturbance rejection capacity of the TCV X-Point-Target and its overall detachment behaviour from a dynamics perspective on control-relevant timescales. We systematically introduce sinusoidal perturbations in the XPT configuration for three different scenarios: 
\begin{enumerate}
    \item{D$_2$ fuelling gas valve modulations, raising the core density}
    \item{N$_2$ impurity seeding via gas valve modulations in high power conditions while keeping the core density constant}
    \item{Heating perturbations via electron cyclotron resonance heating (ECRH) in the core plasma at high power conditions, whilst keeping the core density and impurity injection constant }
\end{enumerate}
These experiments are benchmarked to a long-legged conventional divertor (Single Null - SN) configuration with similar core shape, plasma scenarios, and perturbations. It is demonstrated that the XPT divertor features an inherent disturbance rejection capacity to fuelling, seeding and ECRH heating modulations at its secondary X-point, for perturbation frequencies spanning $f=$ 4-45 Hz. This suggests that the reduction of detachment sensitivity at the secondary X-point observed in quasi-steady state \cite{Lee2025XPT} extrapolates towards the dynamic regime as enhanced disturbance rejection. In the region upstream of the secondary X-point, the dynamic response of the detached state in an XPT and SN divertor is similar. \par
This work starts with an overview of the experimental setup and system identification method in Section \ref{sect: Experimental setup and methods}. Section \ref{sect: D2 fuelling} presents the results of the D$2$ fuelling perturbation experiments. Section \ref{sec: High power scenarios} extends this analysis to high power N$2$ seeding and ECRH power perturbation experiments, examines the physical origin of the observed disturbance rejection, and discusses its implications for future reactors.

\section{Experimental setup and methods}\label{sect: Experimental setup and methods}
All experiments presented in this work were conducted in the Tokamak à Configuration Variable (TCV) \cite{Theiler2026TCVOverview}, a medium sized tokamak with major radius $R_0 = 0.88$ m, minor radius $a = 0.25$ m, and a carbon first wall. TCV is equipped with 16 independent poloidal field coils, allowing for precise core and divertor shape control \cite{Mele2025_ShapeControl}. This capability enables the formation of the XPT divertor \cite{LaBombard2015XPT} on TCV \cite{Theiler2017_FirstADCs,Raj2022SXPT,Lee2025XPT}, and facilitates the design of a similar core shape and plasma scenario for both XPT and SN divertors. Thereby, a comparison of the detachment dynamics in both divertors is enabled while they operate in similar scenarios, as described in Section \ref{subsec: Plasma scenarios}. An overview of the diagnostics that were employed to measure the evolution of the detached state is provided in Section \ref{subsec: Diagnostic setup}. Section \ref{subsec: System identification} details the system identification approach that combines diagnostic outputs with actuator inputs to identify detachment dynamics.

\begin{figure}[t]
    \centering
    \includegraphics[width=1\linewidth]{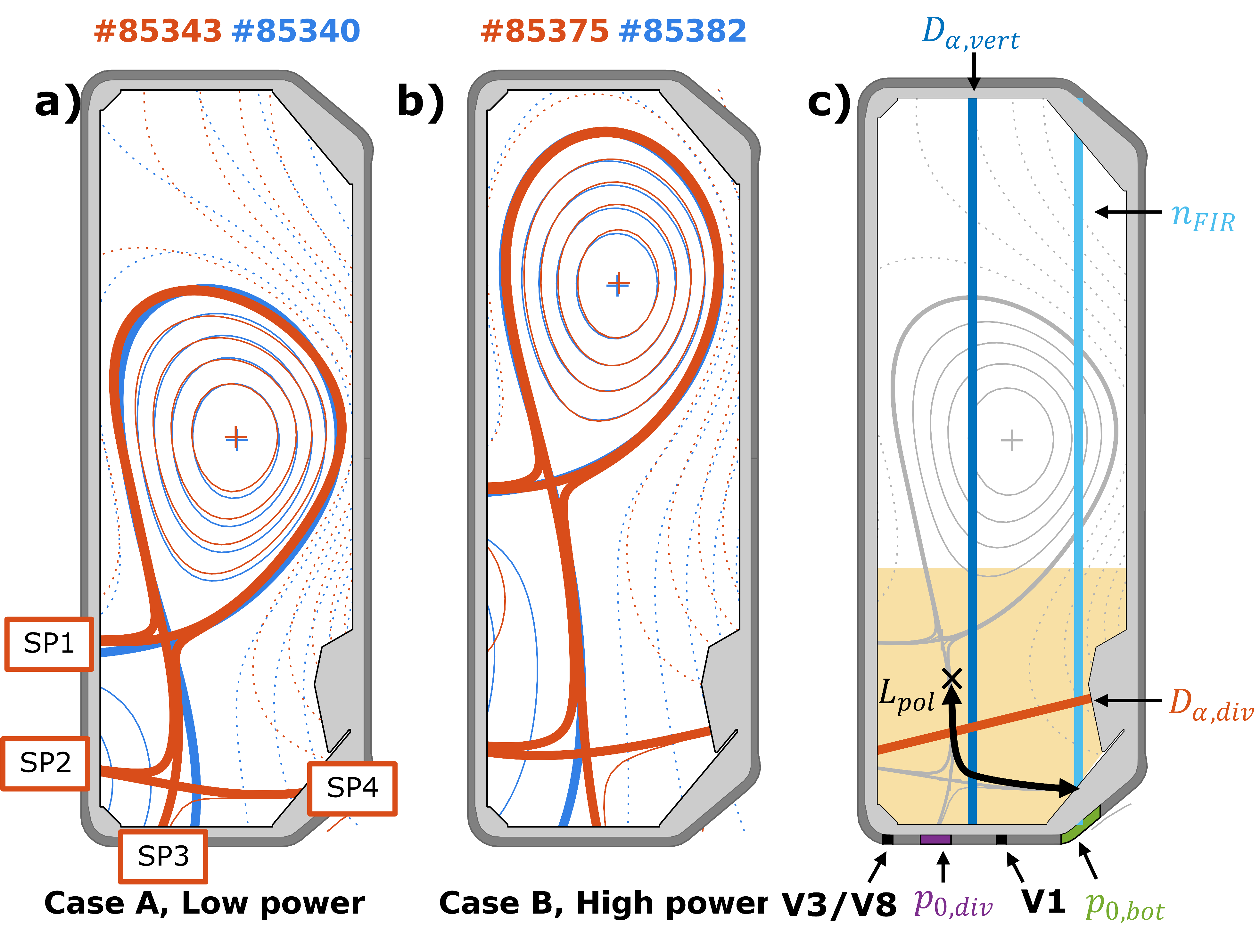}
    \caption{Magnetic equilibria and diagnostic locations. a) and b) show the magnetic equilibrium reconstructions at $t=1.2$ s for the SN (blue) and XPT (red) configurations for two different plasma scenarios; Case A: Ohmic heating, and Case B: Ohmic + ECRH. c) Diagnostic overview including the Multispectral Advanced Narrowband Tokamak Imaging System (MANTIS) in orange \cite{Perek2019MANTIS}, $D_{\alpha, vert}$ and $D_{\alpha,div}$ for line-integrated Balmer-$\alpha$ emission intensity via photodiodes in red and dark blue, line-integrated core density $n_{FIR,2}$ via interferometer chord two (FIR \cite{Barry1999FIR}) in light blue, neutral pressure measurements $p_{0,div}$ and $p_{0,bot}$ via ASDEX pressure gauges (APG \cite{Sun2025APGinTCV}) in purple and green.  }
    \label{fig: experimental setup}
\end{figure} 

\subsection{Plasma scenarios}\label{subsec: Plasma scenarios}
Two different operational scenarios are chosen to study the disturbance rejection capability of the detached state in both SN and XPT configurations under increasingly demanding conditions for detached operation. First, a low power Ohmic scenario is considered as Case A in which major benefits were found for detached operation in quasi-steady state for the XPT divertor by Lee et al. \cite{Lee2025XPT}. This scenario operates with $I_p = 300$ kA, a toroidal field strength of $B_t = 1.4$ T in reversed toroidal field configuration, which is unfavourable for H-mode \cite{Wagner1982Hmode} access. The corresponding magnetic equilibria are shown in Figure \ref{fig: experimental setup}(a), with the XPT and SN configurations in red and blue respectively. In this scenario, D$_{2}$ fuelling perturbations were introduced via Valve 1 (V1 in Figure \ref{fig: experimental setup}), without gas injection from other valves. An overview of the relevant discharges, including parameters of the perturbation signals is provided in Table \ref{tab: fuelling shotoverview}. Note that in this scenario, no auxiliary ECRH is added, as core density fluctuations induced by fueling perturbations may impact the ECRH coupling, thus complicating the analysis of fuelling impact on detachment. \par

Secondly, the dynamics of the detached state were examined in a novel high power, low upstream separatrix density, scenario. This scenario approaches the power exhaust challenge expected for SPARC \cite{Lee2026HighPowerScenarios}, based on the normalized parameter $f_{z} \sim \frac{P_{SOL} B}{R}/n_{e,sep}^2$ as a metric for detachment accessibility \cite{Body2024LengyelMetric}, expressed in terms of the impurity fraction $f_z$, power entering the SOL $P_{SOL}$, and upstream electron density in the separatrix $n_{e,sep}$. This Case B scenario \cite{Lee2026HighPowerScenarios} modifies the low power case by employing long-legged SN (blue, 85382) and XPT (red, 85375) configurations shown in Figure \ref{fig: experimental setup}(b) with auxiliary ECRH heating on-axis via the second harmonic X-Mode (X2), thereby enabling the introduction of heating perturbations via ECRH. A maximum of $P_{ECRH} = 520$ kW for the SN, and $P_{ECRH} = 2570$ kW for the XPT is injected. In these high power conditions, N$_{2}$ seeding is required \cite{Lee2026_InPrep} to obtain a detached state, thereby also enabling the introduction of N$_2$ seeding perturbations. N$2$ impurities are injected through Valve 3 or Valve 8, V3/V8 in Figure \ref{fig: experimental setup}, which share the same poloidal position but differ in toroidal position. All experiments containing N$_2$ perturbations use V3, as this valve contains a flow meter close to the vessel wall, allowing to distinguish between the dynamics of the gas system and plasma interaction, Section \ref{subsec: System identification}. To estimate a similar level of detachment between the SN and XPT configurations, we use the Detachment Location Sensitivity (DLS) model \cite{Lipschultz2016_DLS1,Myatra2023DLS}, \ref{appendix: DLS}. This reduced model defines a \textit{Detachment driver} $D = n_{e,u} f_z^{1/2}/q_{\parallel,u}^{5/7}$ as a function of the relative detachment front position on the divertor leg (e.g. relative to the X-point) and magnetic geometry in the divertor region, from which the operational space of detachment can be extracted as a function of the upstream density $n_{e,u}$, impurity fraction $f_z = n_z/n_e$, and upstream parallel heat flux $q_{\parallel,u}$. The metric $D$ as a function of front position was used to scale the levels of ECRH and N$_2$ seeding between both XPT and SN configurations. An overview of the relevant discharges is provided in Tables \ref{tab: seeding shotoverview} and \ref{tab: heating shotoverview}. \par


In both scenarios, the XPT configuration is formed by introducing a secondary poloidal magnetic null in the outer divertor leg of a SN configuration. This results in a split of the outer leg into two distinct active branches and multiple strike points (SP in the figure). Here, SP3 is magnetically disconnected, whereas the division of heat and particle fluxes between SP2 and SP4 mainly depends on the radial distance between the separatrices at the outboard midplane in the direction of the far-SOL, denoted as $dR_{u,x2}$, and the toroidal field direction \cite{Carpita2026_InPrep, Lee2026_InPrep}. For all experiments in this work, $dR_{u,x2} = 1.0 \pm 0.5$ mm, which is small relative to the e-folding heat flux width $\lambda_q \approx 4$ mm of the SOL, measured in an Ohmic scenario \cite{Lee2025XPT}. As a result, the majority of the heat flux is expected to be directed to SP4. \par

\begin{figure*}[ht!]
    \centering
    \includegraphics[width=1.0\linewidth]{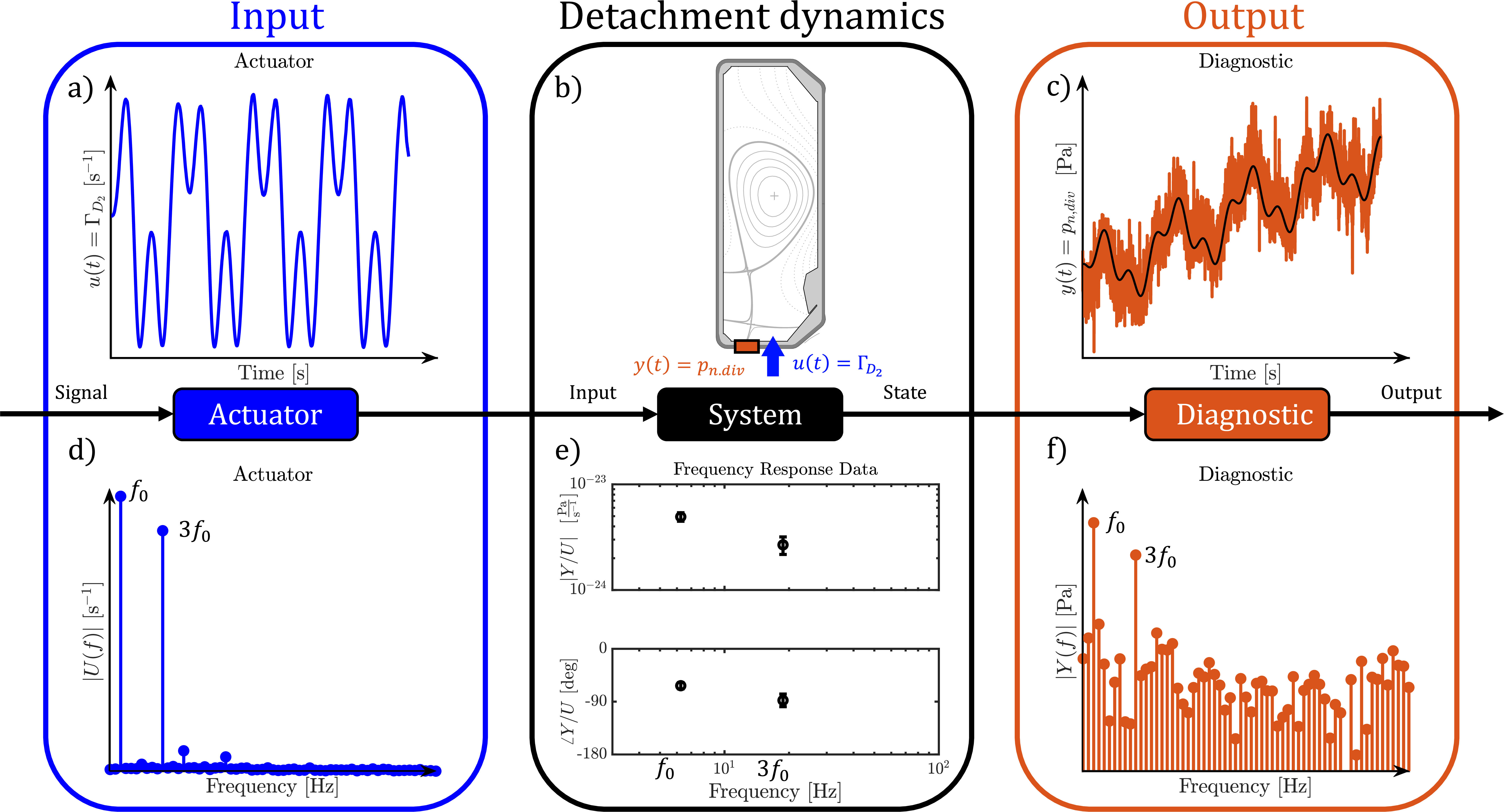}
    \caption{Example of the system identification method to obtain Frequency Response Data (FRD) for TCV discharge $\# 85343$. The actuator (D$_{2}$ gas puff) introduces a multi-sine perturbation input signal $u(t) = \Gamma_{D_{2}} (t)$ (a) to the divertor plasma in (b) with excited frequencies $f_{exc} = f_0\cdot[1,3]$, as shown by the Discrete Fourier Transform (DFT) $U(f) = \mathcal{F}\{u(t)\}$ in (d). As a result, the divertor state evolves, which is measured with various diagnostics. The resulting output (divertor neutral pressure in this case) is shown in (c) with $y(t) = p_{n,div}(t)$ in red, and LPM \cite{Berkel2020LPM} reconstructed response in black. (f) shows the DFT of the output signal $Y(f) = \mathcal{F}\{y(t)\}$. As a result, the linearized dynamics can be identified via the FRD $H(f_{exc}) = Y({f_{exc})/U(f_{exc})}$ in (e) with error bars determined via the LPM \cite{Berkel2020LPM}. }
    \label{fig: SYSID overview}
\end{figure*}

\subsection{Diagnostic setup}\label{subsec: Diagnostic setup}
The diagnostics employed to study detachment dynamics are shown in Figure \ref{fig: experimental setup}(c). This work relies heavily on MANTIS \cite{Perek2019MANTIS} to study detachment dynamics. The lower divertor MANTIS system provides bandpass filtered images of selected spectral lines with an approximate view that corresponds to the transparent orange region in Figure \ref{fig: experimental setup}. After tomographic inversion \cite{Perek2021InversionMANTIS}, a poloidal (2D) map of this emission is obtained. We combine inverted images with LIUQE \cite{Moret2015LIUQE} magnetic reconstruction to track the emission front of the CIII (465 nm) transition along the divertor leg in the poloidal plane, representing a cold ($T_e=5-8$ eV \cite{Martinelli2023_CIII,koenders_systematic_2022,Fevrier2020N2SeededDetachmentTCV}) plasma region upstream of the detachment front (typically at $T_e<5$ eV \cite{Lipschultz2007DivertorPhysics,Verhaegh2017Detachment,Potzel2014DetachmentTemp}). A version of the tracking algorithm described in \cite{Ravensbergen2020FrontTrack} is used, generalized to be compatible with strongly curved divertor legs, including those of an XPT divertor. The CIII front is defined as the position along the divertor leg where CIII emission has dropped by 50 percent of its maximum value \cite{Theiler2017_FirstADCs,Ravensbergen2021NaturePaper}. Although the CIII front is located upstream of the detachment front, its movement along the divertor leg is commonly used as a proxy for detachment evolution and sensitivity \cite{Theiler2017_FirstADCs, Harrison2017DetachmentEvolution, Lee2025XPT}. The position of the CIII front along the divertor leg is defined as $L_{pol}$, which represents the poloidal distance between the front and target along the magnetic field, as shown in Figure \ref{fig: experimental setup}(c). 
\par
Additionally, far infrared (FIR \cite{Barry1999FIR}) interferometer chord number 2, Figure \ref{fig: experimental setup}(c), provides a measure for the line-integrated core density. This chord was selected because it provides a line-integrated core density measurement that is comparable between the SN and XPT configurations. Conversely, other FIR chords include a significant additional contribution from the plasma density in the divertor region in the XPT case, as confirmed by comparison with Thomson scattering data in both core and divertor domains, not shown. A measurement of the core density by FIR was preferred over measurements via Thomson scattering as the acquisition frequency of the latter (60 Hz) is insufficient to resolve the dynamics investigated in this study.  The neutral pressure $p_0$ in the divertor is monitored by two fast ion gauges (green and purple coloured APG \cite{Sun2025APGinTCV} in Figure \ref{fig: experimental setup}), located at different positions in the divertor region. Furthermore, two photodiodes (PD) with vertical (top-down, in dark blue) and divertor (in red) sight lines measure the Balmer-$\alpha$ emission (656 nm) intensity $D_{\alpha}$, which results from the de-excitation of neutral deuterium atoms \cite{Verhaegh2021DetachmentPlasmaMolecule}. The $D_{\alpha}$ emission is commonly used as a proxy for plasma-neutral interactions. 

\subsection{System identification method}\label{subsec: System identification}
The given set of diagnostics is used to identify detachment dynamics with a system identification method, which has been applied to study various dynamical processes in multiple tokamaks \cite{VanBerkel2025sysIDOverview}. With this method, the divertor plasma (the system, described by non-linear physics) is perturbed around an operational point to obtain a measurement of the local linearized dynamics of the detached state. In this work, the operational point constitutes the CIII front position at the divertor leg as a measure for the status of the detached state. Figure \ref{fig: SYSID overview} provides an overview of the method, which is detailed in this section.\par
After detaching the CIII front from the target, an input perturbation $u(t)$ is introduced via different actuators, i.e. D$_{2}$ gas fuelling, N$_{2}$ gas impurity seeding or ECRH power modulations (Figure \ref{fig: SYSID overview}(a)). The perturbed gas valves (V1,V3) are feedback controlled and contain a flow meter near the vessel wall, such that dynamical effects from neutral transport in the gas system outside of the vessel can be separated from the identified detachment dynamics. The perturbation in the input signal is chosen to be a multi-sine opposed to e.g. a block-wave or step function, as it enables the excitation of a specific set of frequencies (timescales), provides a good signal-to-noise ratio (SNR), and facilitates the differentiation between linear and non-linear behaviour \cite{VanBerkel2025sysIDOverview}. A detailed description of the multi-sine signal design is given in \ref{appendix: Signal design}. \par
Upon actuation with carefully chosen frequencies $f_{exc}$ in the multi-sine signal, a response in various plasma parameters can be observed, e.g. in divertor neutral pressure, Figure \ref{fig: SYSID overview}(c). We analyze these input and output signals to infer Frequency Response Data (FRD), as visualized in a Bode plot in Figure \ref{fig: SYSID overview}(e). The gain in the top of the Figure quantifies the magnitude with which the output responds to the input, whereas the phase lag at the bottom characterizes the time delay of the output relative to the input, both across a range of frequencies (timescales). 

In practice, the frequency response is obtained by applying a Discrete Fourier Transform (DFT) to both the input, $u(t) \xrightarrow{\mathscr{F}} U(f)$, and the output, $y(t) \xrightarrow{\mathscr{F}} Y(f)$, Figure 2(d,f). Dividing the Fourier components at the excited frequencies,
$H(f_{exc}) = Y(f_{exc})/U(f_{exc})$, yields the Frequency Response Data (FRD)
points, Figure \ref{fig: SYSID overview}(e). These are subsequently corrected with the Local Polynomial Method (LPM) [41] for distortions in the DFT caused by slow transients (originating from e.g. pump dynamics or wall effects) and drifts of the operational point. This methodology follows the procedure described by Van Berkel et al. [22] and is detailed in Appendix A.2. The LPM-corrected FRD points are visualized in a Bode plot, Figure 2(e), which displays the gain $|H(f_{exc})|$ and the phase lag $\angle H(f_{exc})$ as a function of frequency.


Altogether, the FRD provide a measurement of linearized dynamics that does not require an \textit{a priori} physics model, which we use to compare detachment dynamics between SN and XPT discharges.

\begin{figure*}[ht]
    \centering
    \includegraphics[width=\textwidth]{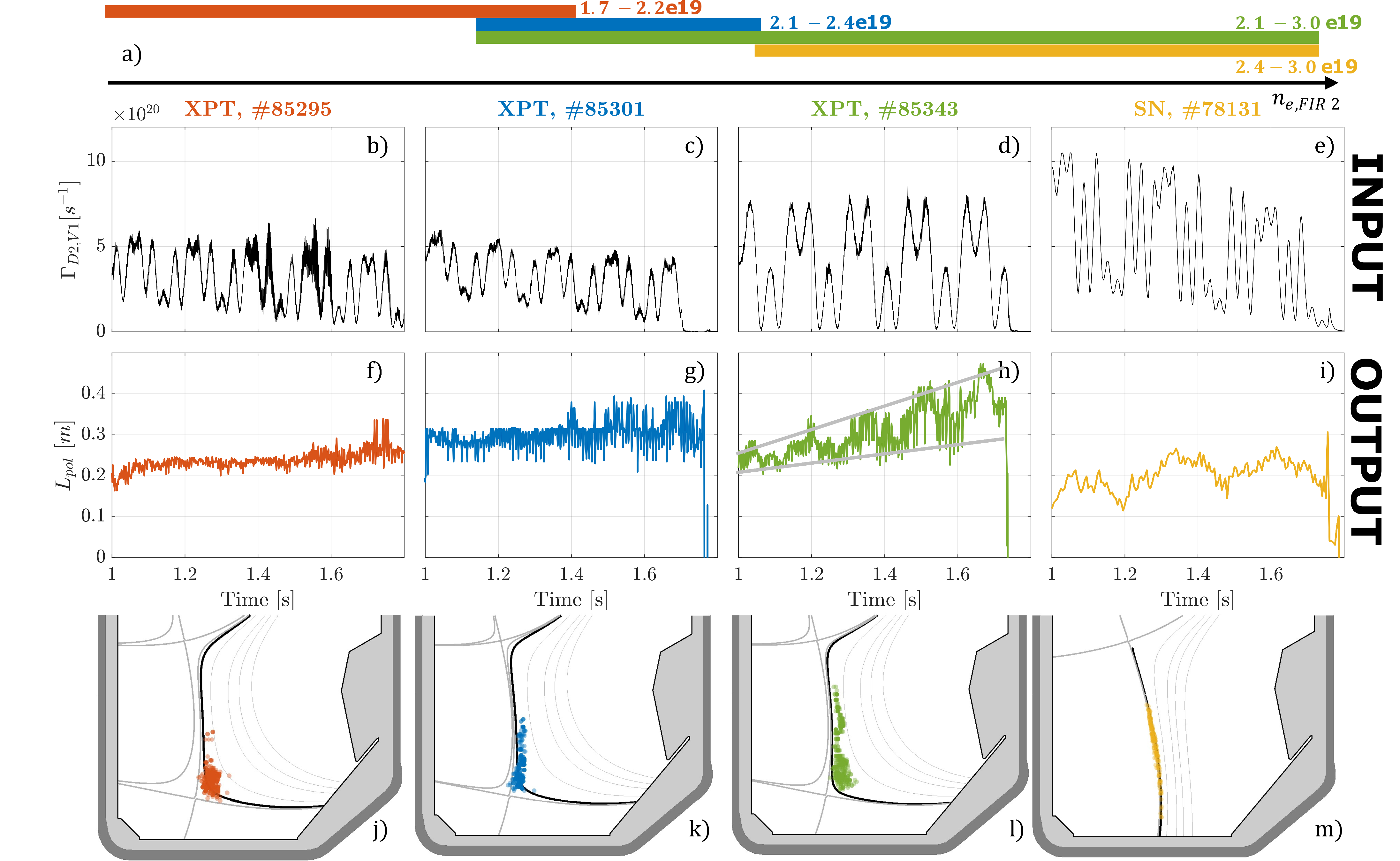}
    \caption{Comparison of the CIII front response in time domain at different operational points in discharges 78131 (SN) and 85295, 85301, 85343 (XPT). (a) shows the line-integrated core density $n_{e,FIR 2}$ of these discharges at the begin and end of the perturbative phase. Multi-sine gas inputs (b-e) and the resulting CIII front movement outputs (f-i) are shown for all discharges. (j-m) shows the poloidal projection of the detected CIII fronts. The black curves represent the flux tube along which the fronts are tracked. The grey lines in (i) indicate an increasing amplitude of the output response for constant input amplitude, suggesting enhanced disturbance rejection in the region close to the secondary X-point.}
    \label{fig: FrontMovementInEquilFuelling}
\end{figure*}


\begin{figure}[h]
    \centering
    \includegraphics[width=1\linewidth]{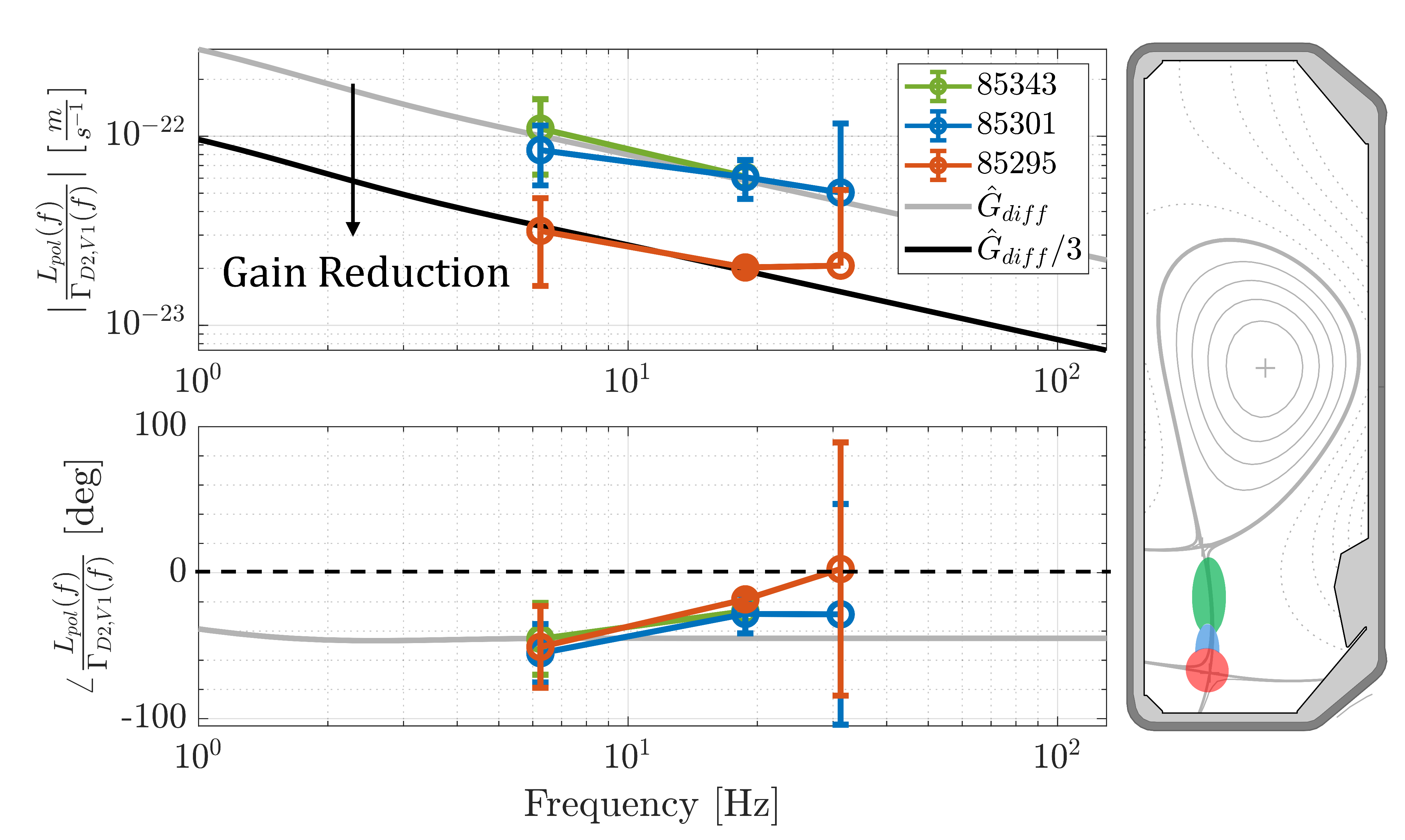}
    \caption{FRD for XPT discharges 85295, 85301 and 85343, with D$_2$ gas puff input and CIII front $L_{pol}$ output. As the CIII front approaches the secondary X point in 85295, a significant reduction in gain is observed compared to the upstream region in 85301 and 85343. Fitting the diffusion based model $G_{diff}$ defined by Koenders et al. \cite{koenders_systematic_2022}, a gain reduction of 3 is measured at the secondary X-point. Since the model $G_{diff}$ is based on SN discharges, it is indirectly a way to compare the SN and XPT geometries, showing that the region upstream of the secondary null (e.g. 85301, 85343) has identical behaviour compared to the SN geometry. The SNR of the FRD in discharge 85295 is poor as shown in Table \ref{tab: SNR table fuelling shots}. Please note that a phase response larger than 0 is considered non-physical as that would implicate a response in $L_{pol}$ before D$_2$ gas is injected.}
    \label{fig: FRF XPT per region}
\end{figure}

\begin{figure}
    \centering
    \includegraphics[width=1\linewidth]{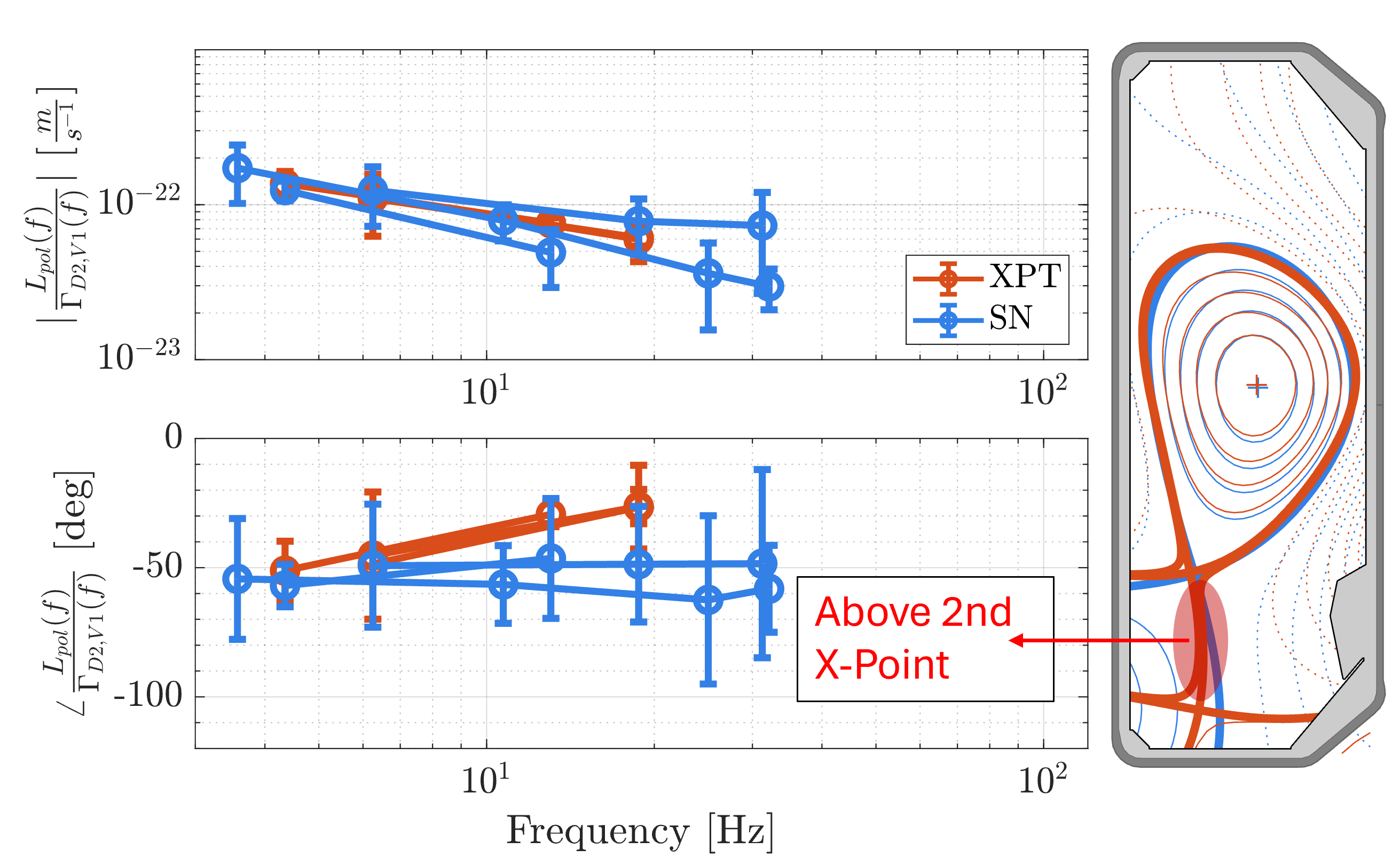}
    \caption{FRD of SN discharges 78131, 85340, and 85347 (blue) and XPT discharges 85343, 85350, and 85353 (red). The CIII front in all XPT discharges drifts from the secondary X-Point towards the first X-Point, indicated with the red transparent region. FRD from the same shot are connected with a line. }
    \label{fig:FRFFuellingSNvsXPT}
\end{figure}

\section{Detachment front dynamics under D$_{2}$ fuelling perturbations}\label{sect: D2 fuelling}
In this section we consider the results of the D$_{2}$ fuelling perturbative experiments in both XPT and SN configurations. The scenarios described by case A in Section \ref{subsec: Plasma scenarios} were used for this study. We first analyse and compare CIII front dynamics in both SN and XPT discharges. This is followed by a comparison of the overall dynamics between the SN and XPT, as measured by multiple diagnostics. \par

\begin{figure*}[ht]
    \centering
    \includegraphics[width=\textwidth]{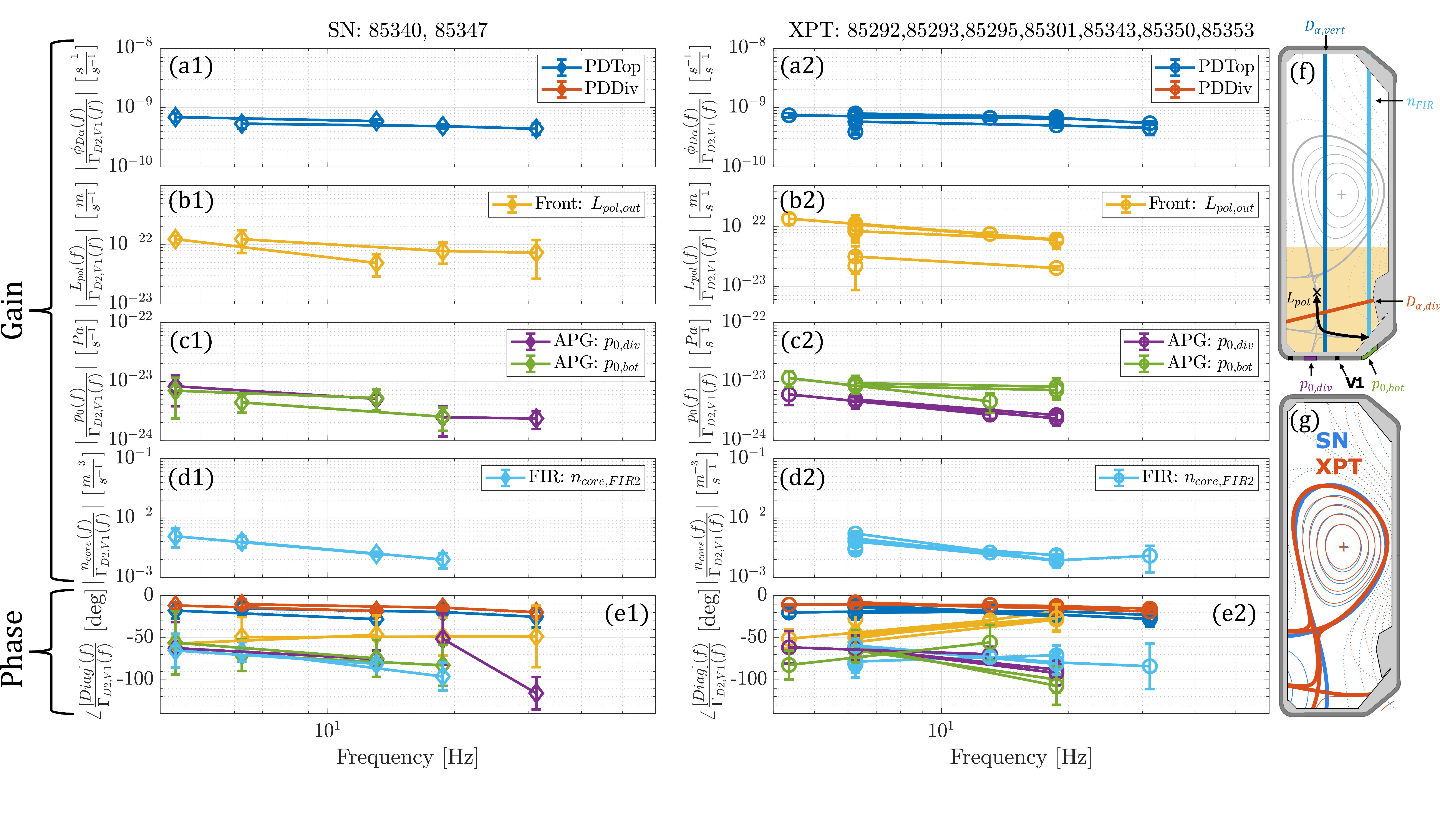}
    \caption{FRD of SN discharges 85340 and 85347 on the left, and XPT discharges 85292, 85293, 85295, 85301, 85343, 85350, 85353 on the right. The top four plots for both SN and XPT represent the FRD points of different diagnostic outputs D$_{2}$ fuelling perturbation input with varying multi-sine signals. The operating point of the CIII front varies between discharges and spans the full space between target and primary X-point in XPT discharges, showing different gain of its response in b2.}
    \label{fig: MultiOutput}
\end{figure*}

\subsection{XPT and SN front dynamics comparison}\label{subsec: XPT and SN comparison}
Focusing on the first three columns with XPT discharges in Figure \ref{fig: FrontMovementInEquilFuelling}, enhanced disturbance rejection of the CIII front  to fuelling perturbations is observed in the region close to the secondary X-point, Figure \ref{fig: FrontMovementInEquilFuelling}(f,j). Conversely, the response becomes more pronounced, similar to a SN divertor in Figure \ref{fig: FrontMovementInEquilFuelling}(i,m), as the CIII front drifts from the secondary to the primary X-point, Figure \ref{fig: FrontMovementInEquilFuelling}(g,h,k,l), as the fuelling levels and core density are raised prior to the perturbative phase, Figure \ref{fig: FrontMovementInEquilFuelling}(a). This is especially evident in discharge 85343, where the amplitude of the CIII front output increases upon constant input amplitude fuelling perturbations, as it drifts from the secondary towards the primary X-Point, Figure \ref{fig: FrontMovementInEquilFuelling}(h,l). \par
The FRD points corresponding to these three XPT discharges, Figure \ref{fig: FRF XPT per region}, reveal a factor 3 gain reduction to fuelling perturbations at the secondary X-point compared to the upstream region between the secondary and primary X-point. Furthermore, a similar slope in gain attenuation and similar phase response are observed across both regions, indicating similar dynamics that drive the evolution of the CIII front. This is corroborated by the diffusive transport model $\hat{G}_{diff}$, regressed from system identification experiments with SN configurations by Koenders et al. \cite{koenders_systematic_2022}. As shown in Figure \ref{fig: FRF XPT per region}, the FRD points obtained upstream of the secondary X-point overlap with $\hat{G}_{diff}$, whereas those obtained at the secondary X-point overlap with $\hat{G}_{diff}$ scaled down by a factor 3, $\hat{G}_{diff}/3$. \par
Several factors must be considered when interpreting the presented FRD. First, the SNR of the FRD points in discharge 85295 close to the secondary X-point is comparatively poor, as the minimal movement of the CIII front in this region produces a weak output signal relative to the noise floor in the frequency domain, Table \ref{tab: SNR table fuelling shots} and \ref{appendix: Small amplitude perturbations}. This is compounded by the reduced amplitude of the excitation signal in these discharges, which was chosen because larger perturbations would cause the CIII front to move upstream of the secondary X-point. Second, the phase response in all discharges decreases towards higher frequencies, which is unphysical, as a system with a pure time delay can only exhibit a constant or increasing phase lag with frequency. A clear cause for this discrepancy was not identified. Altogether, these limitations imply that the reported factor 3 is an indicative estimate, and, given the noise floor of the measurement, a lower bound for the sensitivity reduction at the secondary X-point. This is consistent with quasi-steady-state experiments \cite{Lee2025XPT}, in which a factor 4 reduction in CIII front sensitivity can be estimated, not shown.\\

In contrast, the CIII front response to fuelling perturbations in the SN discharge, Figure \ref{fig: FrontMovementInEquilFuelling}(e,i,m), qualitatively resembles that of the XPT discharges upstream of the secondary X-point, Figure \ref{fig: FrontMovementInEquilFuelling}(g,h,k,l). This is quantitatively confirmed by the corresponding FRD in Figure \ref{fig:FRFFuellingSNvsXPT}, which are of considerably better quality in this region (Table \ref{tab: SNR table fuelling shots}), benefiting from the larger CIII front response to fuelling perturbations. Comparing these FRD points with SN discharges, the gain of the CIII front response overlaps closely between the two configurations across all excited frequencies, while the phase response falls within a comparable range of 20-50 degrees for both. This is consistent with the earlier overlap between the upstream XPT FRD and $\hat{G}_{diff}$, since the latter was itself regressed from SN discharges. This indicates that, upstream of the secondary X-point, the dynamic response of the CIII front in an XPT is indistinguishable from that of a SN divertor. Such agreement is in line with the magnetic topology, which is similar between the two configurations upstream of the secondary null. 
\par Altogether, these observations show that the CIII front dynamics in the XPT are strongly position-dependent. Upstream of the secondary X-point, the front behaves similarly to the SN configuration, whereas its sensitivity to fuelling perturbations is markedly reduced once it approaches the secondary null region. Despite the limited FRD quality near the secondary X-point, this reduction is consistently observed across all XPT discharges considered here, for perturbations in the range of $f = 4-31$ Hz, providing a first quantitative indication of the disturbance rejection capacity at the secondary X-point. This reduced sensitivity aligns with the reduction in detachment sensitivity previously observed at the secondary X-point in quasi-steady-state experiments \cite{Lee2025XPT}, indicating that this benefit extrapolates from quasi-steady-state towards control-relevant, dynamic timescales. The reduced sensitivity of the CIII front at the secondary X-point is considered highly beneficial for maintaining the detached state, as the disturbance rejection of the secondary X-Point may function as a passive buffer, thereby simplifying the task for control systems to stay within operational constraints.

\subsection{XPT and SN multi-output comparison}\label{subsec: MultiOutput}
Although the evolution of the CIII front discussed in Section \ref{subsec: XPT and SN comparison} provides a proxy for the evolution of the detached state, it captures only part of the full dynamical picture. To establish whether the reduced front sensitivity observed at the secondary X-point reflects a change in the underlying dynamics of the upstream conditions and detached state, or is instead a localized effect specific to the magnetic geometry of the XPT configuration, we extend the system identification analysis to the evolution of divertor neutral pressure, line integrated core density and D$_\alpha$ emission, measured via the diagnostics introduced in Section \ref{subsec: Diagnostic setup}, Figure \ref{fig: MultiOutput}(f). This allows us to compare, between SN and XPT, (i) the chronological order in which different physical quantities respond to fuelling perturbations, via their phase response, and (ii) the strength of that response, via their gain. All data presented in this section was observed to have a dominantly linear response in \ref{appendix: multi output}, thereby justifying the use of Bode plots for this analysis. The resulting FRD for all sensors are shown in Figure \ref{fig: MultiOutput}, with SN discharges and XPT discharges shown separately. Here, only FRD points were considered with an SNR$>$3. Note that the selected SN and XPT discharges together cover CIII front positions
between the target and the primary X-point, and are not perturbed around a
single $L_{pol}$.

Examining the phase response first, Figure \ref{fig: MultiOutput}(e1,e2), a consistent chronological order is observed across all perturbed frequencies in both the SN and XPT configurations. A response is observed first in the D$_\alpha$ emission measured by the photodiodes with a top-down (D$_{\alpha,vert}$) and divertor (D$_{\alpha,div}$) line of sight, with a phase lag between 10$^\circ$ at low frequencies and 30$^\circ$ at higher frequencies, indicating that D$_2$ neutrals begin interacting with the plasma almost instantly upon injection. This is followed by a response of the CIII front, and finally by the neutral pressure at both divertor locations, which responds simultaneously with the core density measured by FIR chord two. This chronological ordering is consistent with observations at MAST-U and DIII-D \cite{VanBerkel2025sysIDOverview}, and, importantly, is unaffected by the change in magnetic geometry between SN and XPT. This indicates that the sequence of processes driving the response of the detached state is the same in both divertor configurations, i.e. the XPT geometry does not alter the underlying dynamical mechanism, which appears to be dominated by neutral particle dynamics in both cases as hypothesized by van Berkel et al. \cite{VanBerkel2025sysIDOverview}.

Turning to the gain response, the dynamic response of the core density measured by FIR chord two is nearly identical between the SN and XPT configurations, Figure \ref{fig: MultiOutput}(d), and this holds irrespective of whether the CIII front in the XPT discharge is located upstream of, or at, the secondary X-point. This is despite the marked reduction in CIII front gain observed between these same two regions, Figure \ref{fig: MultiOutput}(b2). A similar correspondence between SN and XPT is observed for the $D_\alpha$ emission at both photodiode lines of sight, Figure \ref{fig: MultiOutput}(a). The neutral pressure response is more nuanced. In the SN discharges, both neutral pressure sensors respond with identical gain, Figure \ref{fig: MultiOutput}(c1), while in the XPT discharges, the response measured at the bottom sensor is consistently lower than at the sensor near SP4, Figure \ref{fig: MultiOutput}(c2). This asymmetry, absent in the SN configuration, may not be a consequence of the XPT magnetic topology itself, but rather of the local strike-point angle of the divertor leg on the vertical target at SP4 and the absence of divertor baffling. As shown by Fil et al. \cite{Fil2020NeutralTrapping}, the poloidal angle between the separatrix and the target strongly determines whether recycled and reflected neutrals are redirected back towards the separatrix ('vertical-target') or towards the common flux region ('horizontal-target'), independent of the surrounding divertor magnetic geometry. A locally shallower strike-point angle near SP4 may therefore explain the observed gain asymmetry as a target design effect, rather than one imposed by the XPT geometry.

Altogether, these observations show that the dynamic response of the core plasma, as measured by FIR chord two, is essentially indistinguishable between the SN and XPT configurations, regardless of the operating point of the CIII front in the XPT. Since the core density response does not change appreciably as the CIII front approaches the secondary X-point, the pronounced reduction in CIII front sensitivity reported in Section \ref{subsec: XPT and SN comparison} is not attributed to a change in core plasma dynamics. Instead, this strongly suggests that the reduced sensitivity of the CIII front at the secondary X-point originates specifically from the magnetic geometry of the XPT divertor at the secondary X-point, which correlates with a local increase in total connection length and poloidal flux expansion, rather than from a broader change in the dynamics driving detachment.


\section{Discussion}\label{sec: High power scenarios}
In Section \ref{sect: D2 fuelling} we identified a clear reduction in the sensitivity of the CIII front to fuelling perturbations at the secondary X-point of the XPT, and attributed this behaviour to the local magnetic geometry of the secondary X-point. Sections \ref{subsec: N2 seeding} and \ref{subsec: ECRH Heating} examine whether this disturbance rejection capacity persists under more demanding, high power \cite{Lee2026HighPowerScenarios}, discharges that approach the divertor exhaust challenge expected in devices such as SPARC, by introducing heating and seeding perturbations in the scenario described by Case B in Section \ref{subsec: Plasma scenarios}. Interpreting the front dynamics observed in this scenario, however, warrants discussion, as the CIII front may no longer be indicative of the main heat flux channel evolution in the XPT divertor, as discussed in Section \ref{subsec: Discussion high power} and in more detail in \cite{Carpita2026_InPrep, Lee2026_InPrep}. Consequently, the CIII front in these scenarios should be considered a proxy for the impurity radiation front dynamics rather than the dynamics of the full detached state. Section \ref{subsec: Position dependent front dynamics XPT} examines the physical origin of the observed disturbance rejection capacity of the secondary X-point in the XPT configuration. Finally, Section \ref{subsec: Implications future reactors} discusses the implications of the presented observations for exhaust control of an XPT divertor in a reactor-class device.

\subsection{N$_{2}$ seeding perturbations}\label{subsec: N2 seeding}
\begin{figure}[t!]
    \centering
    \includegraphics[width=\linewidth]{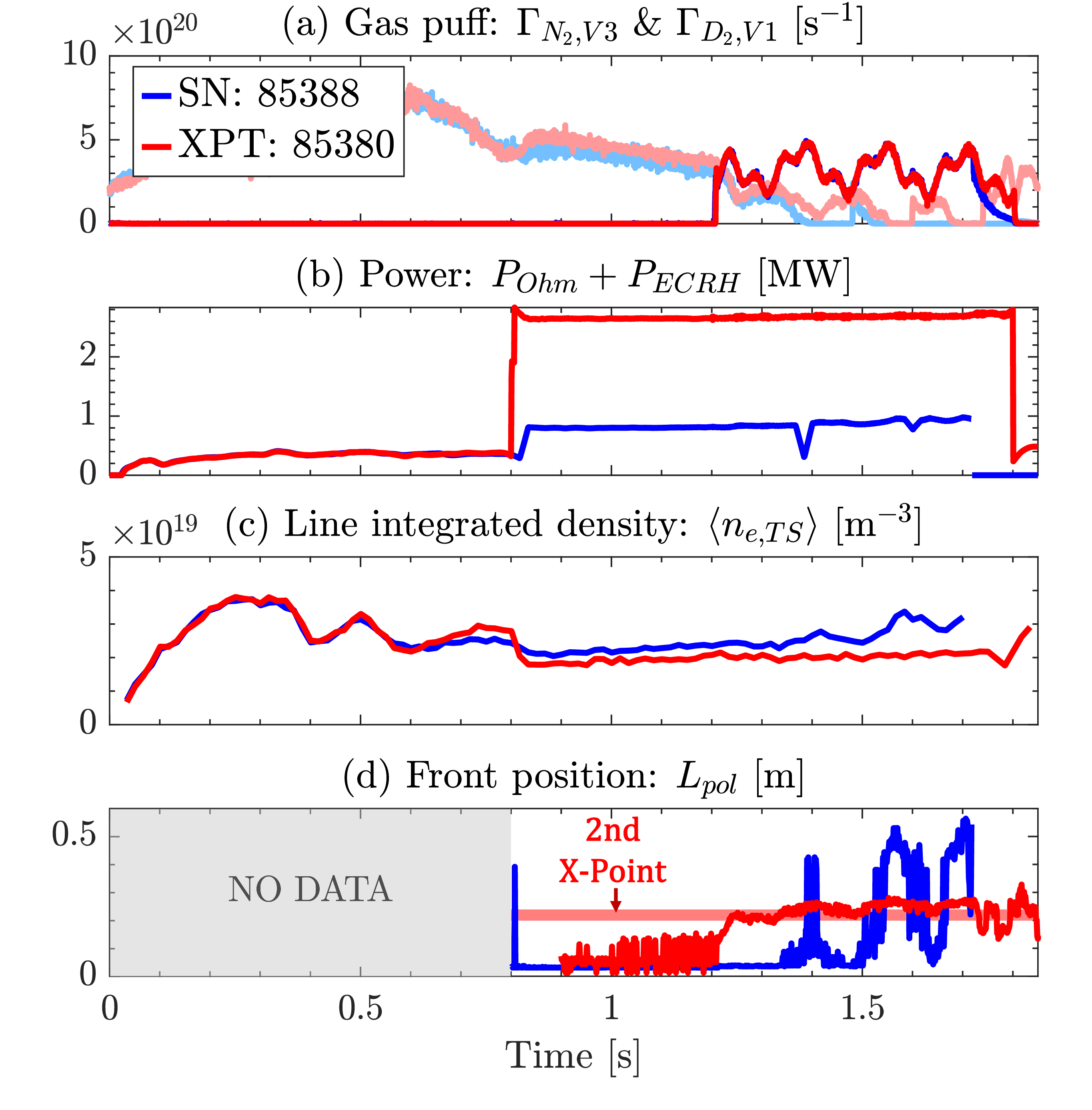}
    \caption{Comparison of N$_{2}$ seeding perturbations in SN discharge 85388 and XPT 85380 in respectively blue and red. Time traces of (a) identical N$_{2}$ gas puff modulations from Valve 3 with $f_{exc} = [6.25, 18.75]$ Hz and background D$_{2}$ gas puff from Valve 1 (light coloured), (b) Ohmic + absorbed ECRH (via TORAY-GA \cite{Matsuda1989TORAY-GA}) power provided to the plasma, (c) Line-integrated core density via Thomson scattering, (d) evolution of the CIII front position. }
    \label{fig: SeedingOverview}
\end{figure}

To further investigate the resilience of the secondary X-point, N$_{2}$ seeding perturbations were introduced in a XPT and SN configuration. Figure \ref{fig: SeedingOverview} shows an overview of the actuator input (a), power conditions (b), line-integrated core density as measured via Thomson scattering (c) and the resulting CIII front evolution output (d) of perturbative N$_{2}$ seeding experiments in SN discharge 85388 and XPT 85380. The N$_{2}$ seeding perturbations were introduced from Valve 3, and are shown in respectively blue and red for the SN and XPT scenarios in Figure \ref{fig: SeedingOverview}(a). The light colored blue and red curves in the same Figure correspond to the D$_{2}$ gas puff from Valve 1. To facilitate the absorption of ECRH power, core density control was enabled during both discharges, in which Valve 1 was used as an actuator. The perturbative phase in these scenarios starts at $t = 1.2$ s. The excitation signal is identical in both scenarios, with $f_{exc} = [6.25, 18.75]$ Hz.  \par

Both discharges have a baseload of ECRH power $P_{ECRH}$ starting at $t=0.8$ s, which differs between the configurations. For similar core density and impurity fraction, the DLS model predicts that the XPT requires more power than the SN to reach the same level of detachment. Thus, to compare the two configurations on an equal footing, the injected ECRH power was scaled accordingly. A total of 2570 kW ECRH power is injected in the XPT discharge, whereas 520 kW of ECRH power is injected in the SN discharge 85388. \par
According to ray-tracing calculations with TORAY-GA \cite{Matsuda1989TORAY-GA}, the absorbed fraction of ECRH power in the SN discharge temporarily drops at $t = 1.38$ s, and $t = 1.6$ s, Figure \ref{fig: SeedingOverview}(c). Here, the first drop in absorbed power is attributed to an outlying Thomson Scattering measurement at that time instance, and is thus not associated with a loss of heating power. The second drop originates from a partial reflection of ECRH power due to a density cutoff. Due to the auxiliary ECRH power, the temperature in the plasma increases, which results in a lower plasma resistivity. Consequently, the coupled ohmic power in the plasma is reduced for the XPT compared to SN. On the contrary, the penetration of N$_{2}$ in the core plasma, Figure \ref{subsec: N2 seeding}(c), throughout the perturbative phase increases the coupled Ohmic power, as $P_{Ohm}\sim Z_{eff}$, Figure \ref{subsec: N2 seeding}(b). Altogether, the total amount of power coupled into the plasma is larger for the XPT (2687 kW on average) compared to the SN (888 kW on average) during the perturbative phase.\par
\begin{figure*}[h!]
    \centering
    \includegraphics[width=\textwidth]{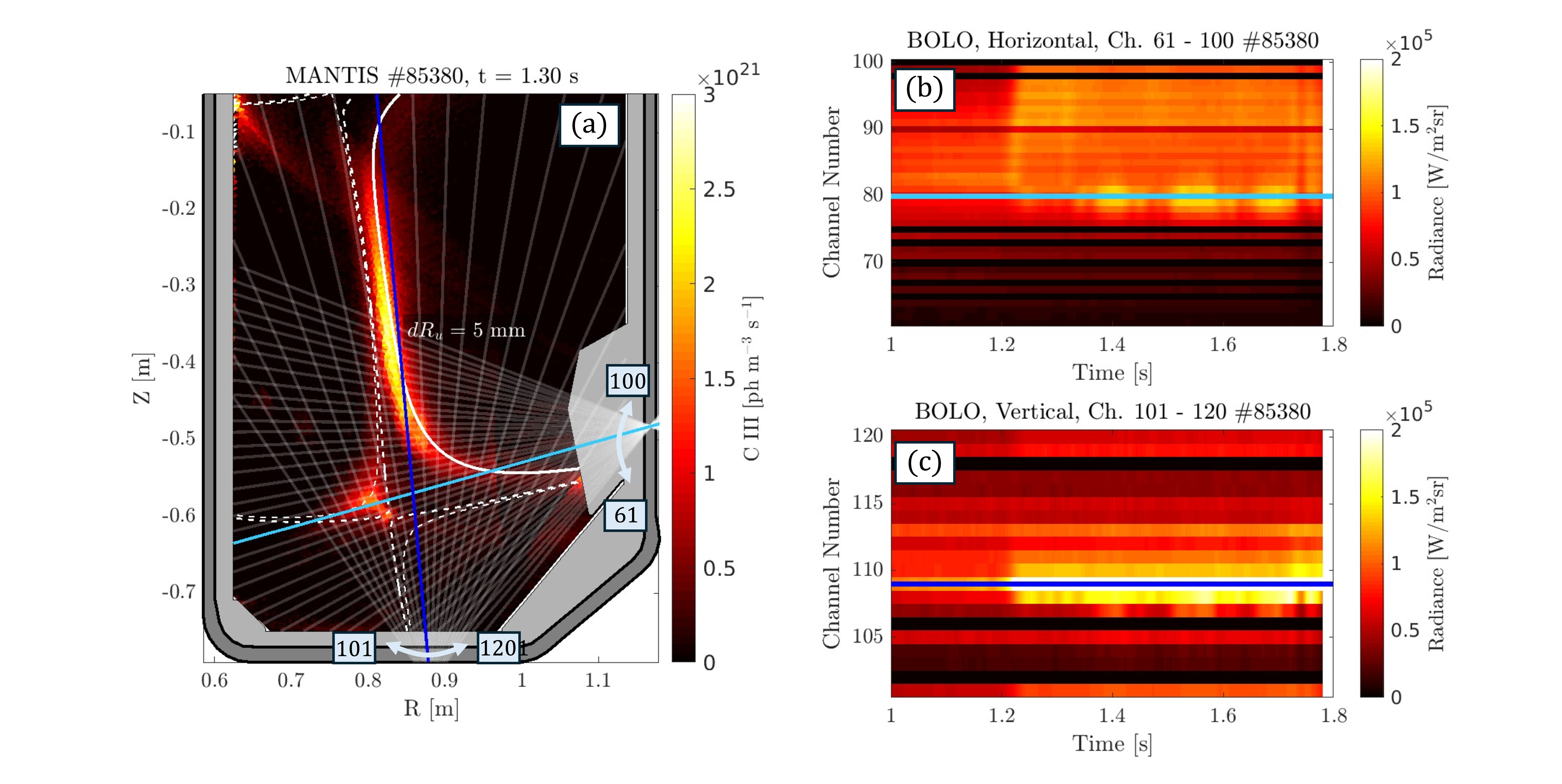}
    \caption{Tomographic inverted image \cite{Perek2021InversionMANTIS} of CIII emission obtained via MANTIS \cite{Perek2019MANTIS} with LIUQE \cite{Moret2015LIUQE} reconstructed equilibrium in white (a), showing two distinct regions with CIII emission in the divertor leg for the XPT. The dotted lines represent the separatrix, whereas the solid white line represents the flux surface at $dR_{u} = 5$ mm. The radiance measured by bolometer \cite{Sheikh2022Radcam} channels 61-100 (b) and 101-120 (c) in transparent grey show that the radiation fronts at channels 80 (light blue) and 109 (dark blue) are not moving upon N$_2$ perturbations. Malfunctioning bolometry channels without data appear as black lines in (b) and (c).}
    \label{fig: MANTISBOLO}
\end{figure*}

 
Despite nearly three times higher coupled power and comparable seeding levels in the XPT compared to the SN, the CIII radiation front in the XPT discharge remains detached and nearly stationary at the secondary X-point throughout the entire perturbative phase, Figure \ref{fig: SeedingOverview}(d). This is corroborated by line integrated radiated power measurements by foil bolometry \cite{Sheikh2022Radcam}, Figure \ref{fig: MANTISBOLO}, which show no significant movement of the radiating region upon N$_2$ seeding perturbations in the XPT discharge. In contrast, the CIII front in the SN is attached in the first part of the perturbative phase, after which it moves back and forth from the target to the first X-point, Figure \ref{fig: SeedingOverview}(d). Note that the response of the front in the SN discharge at $t = 1.6$ s may be significantly affected by the temporary ECRH power cut-off, which could exaggerate the movement of the front throughout the discharge. Furthermore, the penetration of nitrogen into the core plasma significantly affects the upstream conditions by increasing the core density compared to the XPT discharge, Figure \ref{fig: SeedingOverview}(c). Due to these caveats, a comparison between the FRD of both discharges is considered unreliable. \par
Altogether, the reduced sensitivity of the CIII radiation front upon N$_2$ seeding perturbations in the XPT discharge is consistent with, rather than a direct confirmation of, the disturbance rejection capacity of the XPT at the secondary X-point identified for the detached state in the D$_2$ fuelling experiments of Section \ref{sect: D2 fuelling}.

\begin{figure}[ht!]
    \centering
    \includegraphics[width=\linewidth]{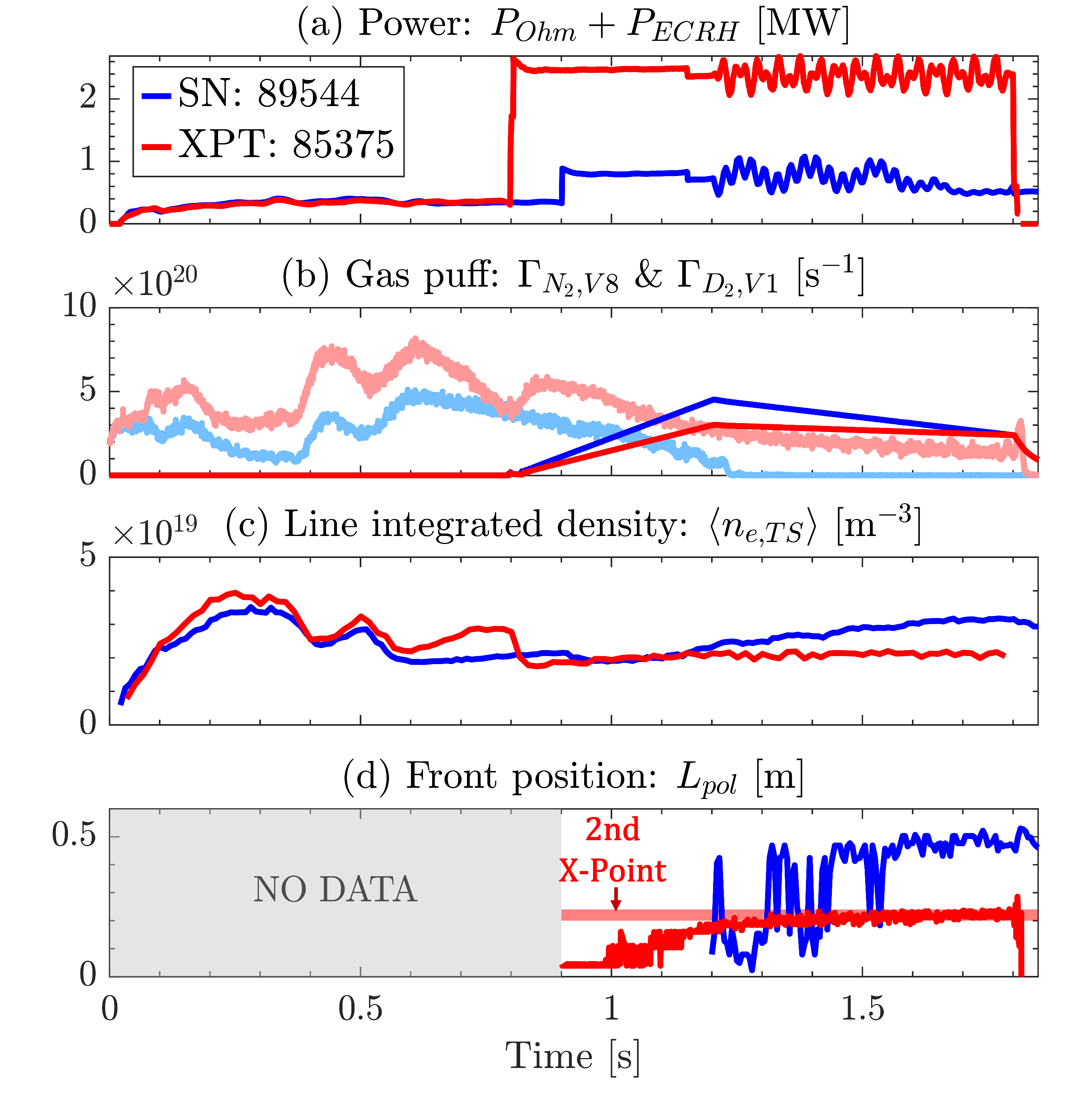}
    \caption{Comparison of ECRH power perturbations in SN discharge 89544 and XPT 85375 in respectively blue and red. Time traces of (a) Ohmic + absorbed ECRH power modulations (according to TORAY-GA \cite{Matsuda1989TORAY-GA}) from G1/L1, with $f_{exc,XPT} = [25.7, 44.6]$ Hz $f_{exc,SN} = [7.7, 38.5]$ Hz , and additional baseload power in the XPT discharge, (b) N$_{2}$ seeding from V8 and D$_{2}$ fuelling from V1 (light coloured), (c) line integrated core density via Thomson scattering, (d) evolution of the CIII front position.}
    \label{fig: HeatingOverview}
\end{figure}

\begin{figure}[h]
    \centering
    \includegraphics[width=1.0\linewidth]{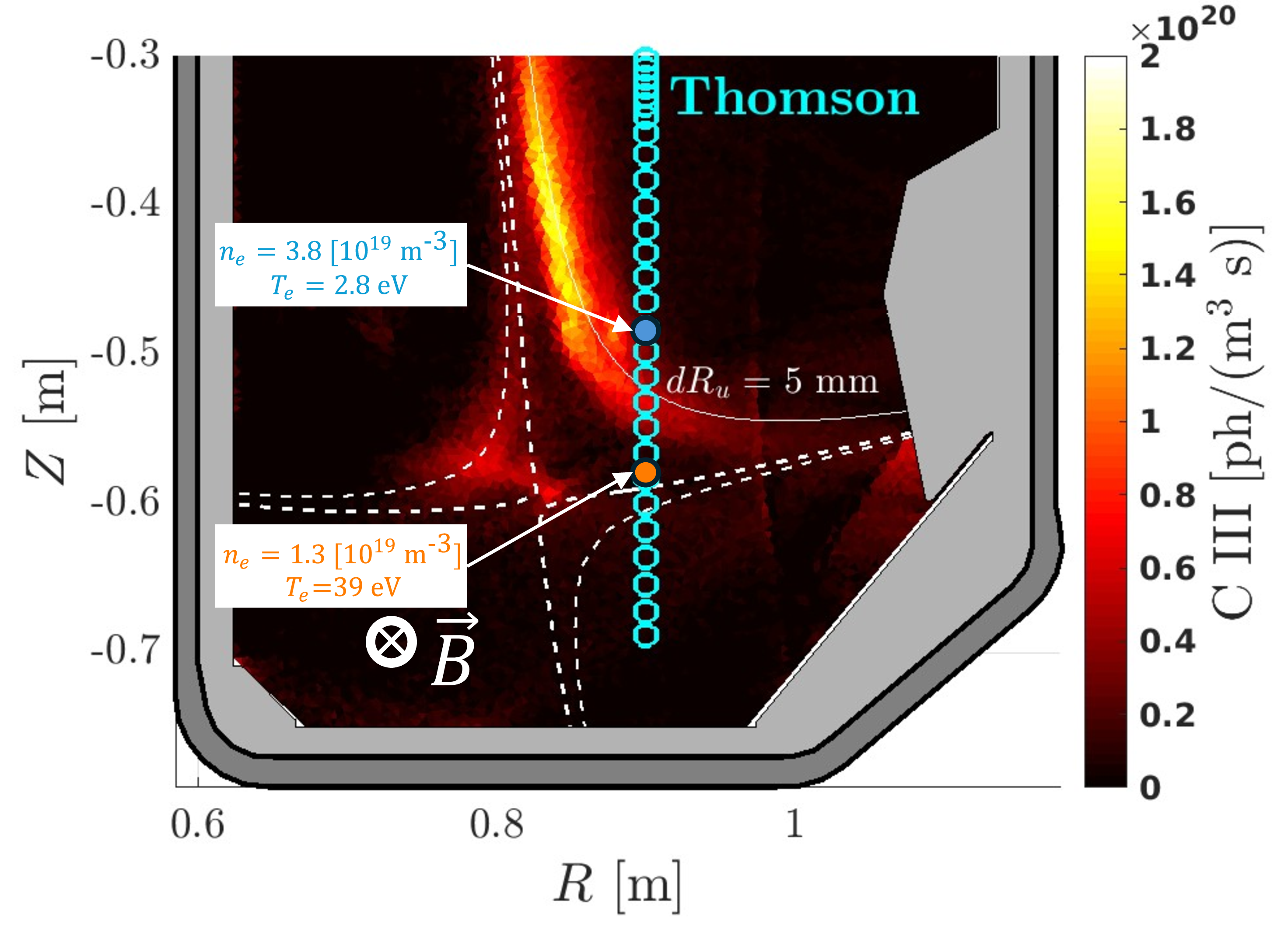}
    \caption{TCV discharge 85375 with an XPT divertor during ECRH modulations in a high power L-mode scenario at $t =  1.4$ s. Two regions of CIII emission are observed in the tomographic inverted image \cite{Perek2021InversionMANTIS} obtained via MANTIS \cite{Perek2019MANTIS} with LIUQE \cite{Moret2015LIUQE} reconstructed separatrix (dashed) in white, along with the flux surface at $dR_{u} = 5$ mm. Thomson Scattering \cite{Blanchard2019TS} measurements of the electron temperature $T_e$ and density $n_e$ at locations in cyan, show a peak in electron temperature close to the separatrix whereas the density peaks in the far-SOL, indicative of cross-field particle transport into the far-SOL.}
    \label{fig: TSandMANTIS}
\end{figure}

\subsection{ECRH power perturbations}\label{subsec: ECRH Heating}
ECRH power modulations were introduced in similar scenarios as the previous section. Figure \ref{fig: HeatingOverview} shows a comparison of XPT 85375 and SN 89544 experiments, displaying the total absorbed ECRH power $P_{ECRH}$ as actuator input (a) along with the coupled Ohmic power, D$_{2}$ fueling (light coloured) and N$_{2}$ seeding gas puffs via respectively Valve 1 and Valve 8 (b), the line integrated core density as measured by Thomson Scattering in (c), and the resulting CIII front evolution $L_{pol}$ in (d), for SN discharge 89544 (blue) and XPT 85375 (red). Here too, core density control was enabled to facilitate the absorbtion of ECRH. Power modulations via ECRH are introduced at $t = 1.2$ s in both SN and XPT scenarios, with $f_{exc,SN} = [7.69, 38.46]$ Hz, and $f_{exc,XPT} = [25.7, 44.6]$ Hz. The amplitude of the power oscillations is 300 kW, along with 1850 kW and 430 kW baseload ECRH power for the XPT and SN respectively. Different base levels of ECRH power were chosen here, similar to Section \ref{subsec: N2 seeding}, to compare the SN and XPT configurations on an equal footing, based on DLS predictions. To detach the divertor plasma, N$_{2}$ seeding gas is injected for both XPT and SN scenarios, where the seeding levels in the SN scenario had to be increased significantly compared to the XPT scenario to detach the CIII front from the target.\par 
The CIII front in the SN responds strongly to the applied power modulations, Figure \ref{fig: HeatingOverview}(d). In contrast, the CIII front in the XPT, tracked along the flux surface at $dR_u = 5$ mm (see Section 4.3), remains remarkably stationary throughout the perturbative phase. Due to the limited response in the XPT discharge, the FRD analysis could not be performed as the response of the CIII front at the excited frequencies in the XPT discharge does not exceed the noise floor. As with the seeding perturbations discussed in Section \ref{subsec: N2 seeding}, this pronounced reduction in the sensitivity of the CIII radiation front - rather than necessarily of the detached state itself - suggests that the disturbance rejection capacity of the secondary X-point extends to power perturbations as well.

\subsection{Discussion on high power results}\label{subsec: Discussion high power}
Sections \ref{subsec: N2 seeding} and \ref{subsec: ECRH Heating} show the CIII front in the XPT to be highly resilient to fluctuations in N$_2$ seeding and ECRH heating compared to the SN configuration, consistent with the disturbance rejection capacity identified for D$_2$ fuelling perturbations in Section \ref{sect: D2 fuelling}. However, this apparent consistency must be interpreted with caution, as the CIII front may not be a good proxy for the evolution of the main heat flux channel in these high power scenarios \cite{Lee2026_InPrep}.\par
To illustrate this, a split in CIII emission is observed in the outer leg of the XPT geometry, Figure \ref{fig: TSandMANTIS}, with one region along the separatrix close to the secondary X-point, and a second in the far-SOL along $dR_u = 5$ mm. In the gap between these two regions, Thomson scattering measures an electron temperature of $T_e = 39$ eV (orange dot in Figure \ref{fig: TSandMANTIS}). At this temperature, C$^{2+}$ ions are further ionised, which depletes the C$^{2+}$ population and prevents CIII emission. In addition, the $n_e$ profile is shifted towards the far-SOL with respect to the $T_e$ profile, such that the electron density peaks at $dR_u >5$ mm, where the temperature is only $T_e = 2.8$ eV (blue dot in Figure \ref{fig: TSandMANTIS}). This shift may cause, or further exacerbate, the gap in CIII emission, and could be associated with strong $\vec{E}\times\vec{B}$ drifts at the secondary X-point \cite{Carpita2026_InPrep,Lee2026_InPrep}.\par

Taken together, these observations indicate that the CIII front may not be indicative of the detached state of all flux tubes in the SOL for this particular high power scenario. The flux tubes close to the separatrix, which carry the majority of the heat flux, largely evade CIII-based detection due to their elevated temperature. The CIII front does, however, continue to trace the radiation front of the divertor plasma. This is supported by bolometry measurements, e.g. during N$_2$ seeding perturbations in Figure \ref{fig: MANTISBOLO}(b,c), which shows shows that the edge of the radiating region towards SP4, at channels 80 and 114 approximately, coincides with the location of the CIII front in Figure \ref{fig: MANTISBOLO}(a), and remains stationary throughout the N$_2$ seeding perturbations. The CIII front is therefore interpreted here as a proxy for evolution of the radiation front, and not for the detached state as a whole.

The observed asymmetry in $n_e$ and $T_e$ profiles may change significantly under a forward toroidal field, as opposed to the reversed field scenarios used in this study, possibly due to a change of the $\vec{E}\times\vec{B}$ drifts \cite{Lee2026_InPrep}. An assessment of forward-field or double-null XPT scenarios with perturbative experiments is therefore an important next step, particularly for understanding possible implications in reactors planning to operate a double-null XPT configuration, such as ARC \cite{Eich2026_ARC}, where the drift direction differs between the upper and lower divertor. \par

In summary, the high power scenarios considered in this study corroborate a reduced sensitivity of the CIII front at the secondary X-point, consistent with the results from fueling perturbations in Ohmic discharges. However, given the appearance of complex electron density and temperature profiles in the XPT divertor, the reduced sensitivity of the radiating region to both N$_2$ seeding and ECRH perturbations may not imply a reduced sensitivity of the detached state itself in these scenarios. Establishing a robust link to the detached state will require an observable more directly tied to the dominant heat flux channel with sufficient time and spatial resolution to resolve the dynamics of interest. 

\subsection{Enhanced disturbance rejection XPT in relation to quasi-steady-state sensitivity}\label{subsec: Position dependent front dynamics XPT}
Having established that the enhanced disturbance rejection capacity of the CIII front to dynamic perturbations at the secondary X-point extends across D$2$ fuelling, N$2$ seeding and ECRH power perturbations (with caveats for the latter two, Section \ref{subsec: Discussion high power}), we now examine this behaviour in a broader context. The disturbance rejection observed at the secondary X-point for D$_2$ fuelling perturbations, Section \ref{sect: D2 fuelling}, is consistent with the reduced sensitivity of the detached state previously reported in quasi-steady-state Ohmic core density ramp discharges \cite{Lee2025XPT}. This work thus shows that the benefit of this reduced sensitivity extends from a \textit{static} to a \textit{dynamical} regime, as an enhanced disturbance rejection capacity against fuelling perturbations, and potentially against heating and seeding perturbations as well.\par 
This is not unique to the XPT divertor topology, as a similar reduction in detachment sensitivity, attributed to total flux expansion, was observed for the MAST-U Super-X divertor \cite{Verhaegh2025DivertorShapingPowerExhaust}. Here too, a superior disturbance rejection at dynamical timescales was observed \cite{Kool2025NaturePaper}, granting actuators more time to respond to disturbances and thereby enabling real-time power exhaust control that was not possible in MAST-U's conventional divertor .\par
\begin{figure}[h]
    \centering
    \includegraphics[width=\linewidth]{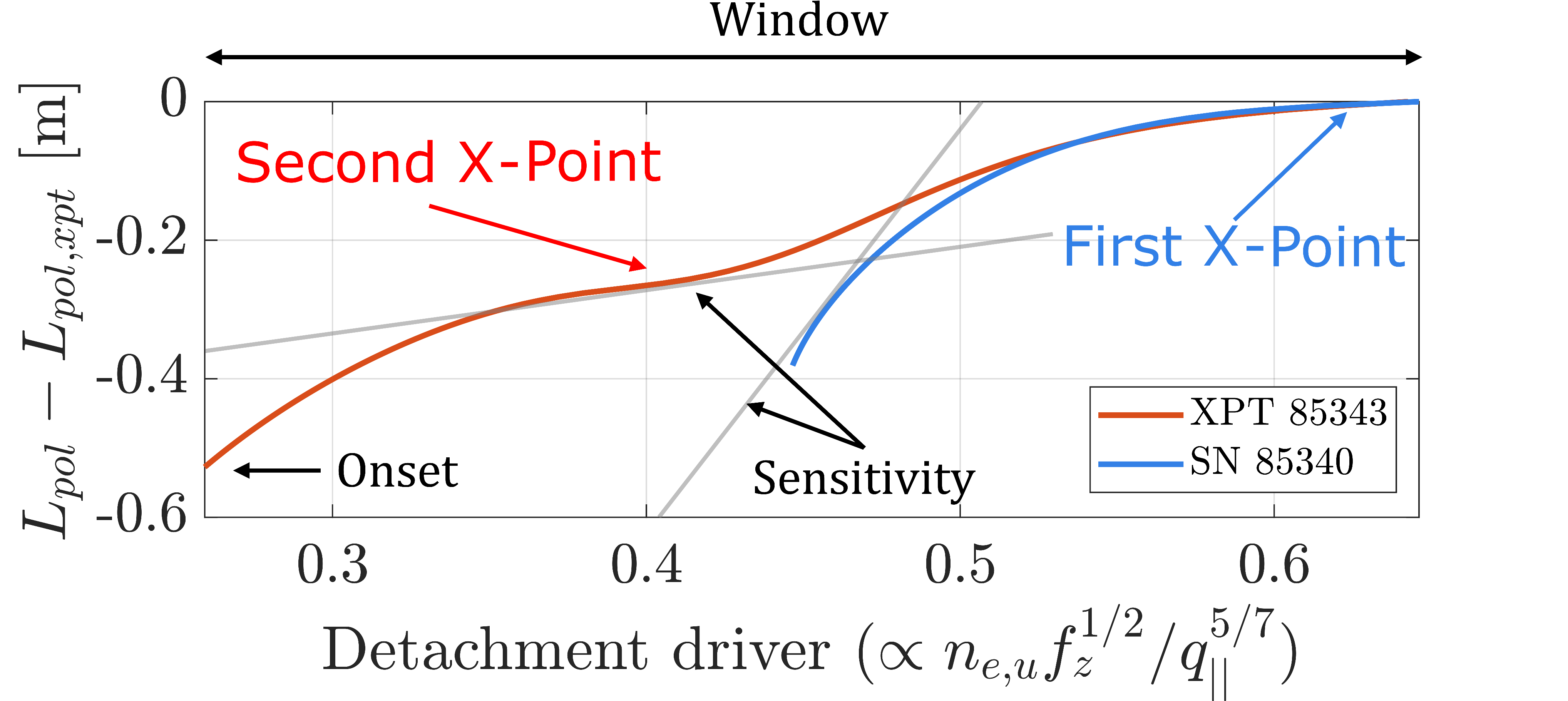}
    \caption{Operational window for detachment of the XPT (red) and SN (blue) as predicted by the DLS model. The poloidal front position is indicated as the difference between the primary X-point $L_{pol,xpt}$ and front position $L_{pol}$, such that the minimum and maximum of the curve correspond to respectively target and primary X-point for both XPT and SN curves. The poloidal locations of the X-points are indicated. The slope of the DLS predicted curves informs on the front sensitivity. As such, the slope of the SN is similar to the slope of the XPT above the secondary X-point, indicating similar sensitivity in that region, consistent with experimental observations. Similarly, the reduced sensitivity at the secondary X-point of the XPT is consistent with experimental observations.}
    \label{fig:DLScomparison}
\end{figure}

One hypothesis for how the XPT configuration achieves this reduced detachment sensitivity is the local increase in connection length near the secondary null. This can be formalized through the Detachment Location Sensitivity (DLS) model \cite{Lipschultz2016_DLS1,Myatra2023DLS}, a reduced, static analytic model that predicts how the detachment onset, sensitivity and window change, relative to a conventional divertor reference, as a function of magnetic geometry (\ref{appendix: DLS}). To do so, the operational space of the detached state is represented by a front position along the divertor leg, $L_{pol}$, and expressed as a function of the upstream electron density $n_{e,u}$, the impurity fraction $f_z$ and the upstream parallel heat flux $q_{_\parallel,u}$, which combines to a lumped parameter $n_{e,u} f_z^{1/2} / q_{\parallel,u}^{5/7}$ (Detachment driver, Figure \ref{fig:DLScomparison}).\par
Taking the magnetic geometries of SN and XPT shapes in Figure \ref{fig: experimental setup}(a) as an input for the DLS model, the secondary X-point reduces the sensitivity to the detachment driver $D$ compared to the SN configuration, Figure \ref{fig:DLScomparison}. Furthermore, since the magnetic geometries of the SN and XPT are similar upstream of the secondary X-point, the DLS model predicts a similar detachment sensitivity there for both configurations, as well as a similar detachment front position relative to the X-point. Both trends in the change in front sensitivity as a function of front location in the XPT geometry are qualitatively consistent with quasi-steady-state XPT experiments in TCV \cite{Lee2025XPT}, MAST-U \cite{Lonigro2026XPTvsSXD}, as well as the dynamic experiments presented in this work. \par
Altogether, these observations show that the reduction of front sensitivity at the secondary X-point, first identified in static and quasi-steady-state analyses \cite{Lee2025XPT}, is qualitatively captured by the DLS model and extrapolates to the dynamic regime probed in this work.

\subsection{Implications for future reactors}\label{subsec: Implications future reactors}
Reactor operation requires exhaust control to maintain the detached state within a viable operating window despite disturbances with ranging timescales, from too fast to actuate on to actionable. This requires considering the dynamical response of the divertor for this range of timescales. Previous work has shown that the X-Point Target \cite{Lee2025XPT} has an increased operating window and a reduced detachment sensitivity in quasi-steady-state. This work shows that not only the sensitivity is reduced in quasi-steady-state ($f=0$ Hz), but for a large range of control-relevant (actionable) timescales ($f=4-45$ Hz). In other words, reduced detachment sensitivity in quasi-steady-state seems correlated with superior disturbance rejection; dynamic perturbations in heating, seeding and fueling impact the radiation region less in the X-Point Target. This is consistent with findings in the Super-X Divertor configuration at MAST-U \cite{Verhaegh2025DivertorShapingPowerExhaust, Kool2025NaturePaper}, Section \ref{subsec: Position dependent front dynamics XPT}. Combined, this suggests that not only the operational window can be increased by certain Alternative Divertor Configurations, their reduced sensitivity in quasi-steady-state makes them less sensitive to dynamic perturbations.\par

This is important for reactor operation, as operation needs to be maintained within a viable operating window throughout these dynamic timescales. Once a viable operating point is reached, this disturbance rejection makes it less likely for heating, seeding and fueling perturbations to exceed operational limits. Especially that this disturbance rejection is maintained for heating and seeding perturbations is an important finding, relevant for reactors, where excursions in heating and seeding are expected due to actuator action (e.g. pellet injection, ramp-up/-down of a discharge), plasma instabilities, and potential off-normal events such as actuator failures, e.g. in auxilliary heating systems \cite{Kool2026_STEP}. These findings provide further confidence of the benefits of an X-Point Target divertor, envisioned in future devices ARC \cite{Eich2026_ARC} and, possibly, SPARC \cite{Creely2020SPARCOverview}.\par 
Yet challenges for power exhaust control in their Double Null (DN) X-Point Target configuration remain. One particular control challenge is that the detachment of the radiation region from the target does not necessarily result in sufficiently low target temperatures, heat fluxes or target detachment, as highlighted in section \ref{subsec: Discussion high power} and discussed in more detail \cite{Carpita2026_InPrep,Lee2026_InPrep}. Therefore, control strategies need to rely on more than a single sensor to accurately assess the detached state, and must be developed for both upper and lower divertors independently, given their different drift behaviour. However, a multi-sensor, multi-actuator model-based control strategy is a common requirement for any divertor configuration and requires specific study and control strategy design for the chosen divertor design \cite{Kool2026_STEP}.\par
A second challenge is that the observables may be sensitive to the precise combination of the secondary X-point location $dR_{x2,u}$ and plasma parameters through $E\times B$ drifts (which are different in lower and upper DN XPT divertors) \cite{Carpita2026_InPrep,Lee2026_InPrep}. As such, integration of magnetic actuation to modify $dR_{x2,u}$ may be essential for exhaust control in an XPT divertor.

\section{Conclusions}\label{sect: conclusions}
In this work we present a dynamic analysis and comparison of experiments with a XPT and SN divertor subjected to perturbations with D$_{2}$ fuelling in Ohmic scenarios, and N$_{2}$ seeding and ECRH power in high power scenarios. The XPT configuration demonstrates a superior disturbance rejection capacity of the CIII-front near its secondary X-point to D$_{2}$ fuelling, N$_{2}$ seeding and ECRH power perturbations, observed in both Ohmic and auxiliary ECRH-powered experiments. For D$2$ fuelling perturbations in Ohmic scenarios, this confirms that the previously observed benefits of reduced detachment sensitivity in quasi-steady-state operation \cite{Lee2025XPT} persist on control-relevant timescales. Furthermore, in the region above the secondary X-point, the dynamic response of the detached state in an XPT is similar to a SN divertor. In ECRH heated discharges, a similar reduction in the sensitivity of the CIII front is observed upon perturbations in N$_2$ seeding and ECRH power, with the caveat that flux tubes carrying the dominant heat flux in these scenarios remained attached. Nevertheless, these results strongly suggest that the observed benefits of enhanced disturbance rejection in the region of the secondary X-point, also extend towards high power scenarios. The observed disturbance rejection is considered highly advantageous for detached operation, as it could provide a passive buffer to mitigate disturbances, effectively providing actuators with more time to respond to them, thereby simplifying the task of maintaining the detached state throughout a discharge cycle.  

\section{Acknowledgements}
This publication is part of the project Balls-to-the-wall: Keeping Hydrogen Burning with Ice Bullets (Project No. 19695) of the NWO Talent Programme VIDI, financed partly by the Dutch Research Council (NWO). DIFFER is part of the institute’s organization of NWO. This work has been carried out within the framework of the EUROfusion Consortium, partially funded by the European Union via the Euratom Research and Training Programme (Grant Agreement No 101052200 — EUROfusion). The Swiss contribution to this work has been funded by the Swiss State Secretariat for Education, Research and Innovation (SERI). Views and opinions expressed are however those of the author(s) only and do not  necessarily reflect those of the European Union, the European Commission or SERI. Neither the European Union nor the European Commission nor SERI can be held responsible for them. This work was supported in part by the Swiss National Science Foundation.

\newpage
\onecolumn

\appendix 


\section{System Identification}
\subsection{Signal design}\label{appendix: Signal design}
The multi-sine actuation signals utilised for system identification were designed to be compatible with the different plasma scenarios described in Section \ref{subsec: Plasma scenarios} as well as the operational constraints of TCV. The main design objective is to obtain reliable FRD while remaining within actuator limits and preserving an operating point (CIII front location) in the detached state. The multi-sine signals of the input signals $u(t)$ were designed using Crest factor ($C = |u_{max}|/u_{rms}$) minimization \cite{Pintelon2012SYSID}. By reducing this peak-to-RMS ratio, we generate a signal with maximized average power for a fixed peak amplitude, thereby improving the SNR.
Each excited frequency should complete at least three full periods within the available perturbation time to enable the estimation of a standard deviation $\sigma$ on the response via the LPM \cite{Berkel2020LPM}. The typical perturbation time in the presented TCV experiments is $T_{pert} = 0.8$ s, which defines a lower frequency limit of $f_{exc} = 3/T_{pert} = 3.75$ Hz. In the perturbative scenarios that include D$_{2}$ and N$_{2}$ gas puff modulations, a high frequency limit is set by the available gas valve that cannot follow gas flow request above 50 Hz \cite{koenders_systematic_2023}. \par
The total amount of injected gas in the fuelling and seeding experiments further constraints the excitation amplitude. Excess fuelling and impurity seeding cool the divertor throughout the perturbative phase, which shifts the CIII front upstream from its original position. This drift alters the operational point of the detached state and leads to non-local measurements. Contrary, input signals with smaller amplitudes result in a worse SNR, compromising the FRD quality. Within these bounds the selected frequencies match those used in previous system identification studies \cite{koenders_systematic_2023,Koenders2023Control,Ravensbergen2021NaturePaper, Derks2025MultiModel} at TCV on SN configurations. This consistency facilitates comparison across plasma regimes, machine configurations, and actuators. The experiments using ECRH actuation similarly utilizes excitation frequencies that match previous studies \cite{VanBerkel2025sysIDOverview}. In this scenario, the hardware does permit excitation frequencies exceeding 50 Hz. 

\subsection{Frequency Response Data}\label{appendix: Frequency Response Data}
Upon actuation with carefully chosen excited frequencies $f_{exc}$ in the multi-sine signal, the detached state evolves, which can be measured as an output $y(t)$ across various diagnostics, Figure \ref{fig: SYSID overview}(c). To characterize the dynamics that determines the input-output relation, both input and output signals are transformed from the time domain, to frequency domain via a Discrete Fourier Transform (DFT), yielding $U(f) = \mathcal{F}\{u(t)\}$ and $Y(f) = \mathcal{F}\{y(t)\}$ (Figure \ref{fig: SYSID overview}.d,f). To correct for distortions in the DFT caused by slow transients (originating from e.g. pump dynamics or wall effects) and drifts of the operational point, the local polynomial method (LPM) \cite{Berkel2020LPM} is used. \par
By combining the input and output with the LPM correction in the frequency domain, a measurement of the local linear dynamics at the excited frequencies can be extracted as Frequency Response Data (FRD, Figure \ref{fig: SYSID overview}(e)) $H(f_{exc})$, defined as

\begin{equation}\label{eq: FRM}
    H(f_{exc}) = \frac{Y(f_{exc})}{U(f_{exc})}. 
\end{equation}

Two criteria have to be met for the FRD to be a reliable measurement of the linearized dynamics that relates input to output. First of all, the system has to respond dominantly linear upon excitation, which can be identified through the output spectrum $Y(f)$. Linear systems show a response only at the excited frequencies $f_{exc}$. Contrary, dominant non-linear systems also generate a response at (low-order) integer multiples of $f_{exc}$ \cite{Berkel2020LPM}. Therefore, to facilitate the detection of dominant nonlinearities, all excitation signals in this work utilize excitation frequencies with odd-order harmonics (e.g. $f_{exc} = f_{0}\cdot[1,3,5,...]$). This choice ensures that any resulting dominant even-order harmonics in $Y(f)$ can be attributed to even-order non-linearities.  \par
Secondly, the SNR of the excited frequencies in the output $Y(f_{exc})$ should be sufficiently high to exceed the noise floor. The SNR of the output signal is defined as SNR$=|Y(f_{exc})|/\sigma_{noise}$, where $\sigma_{noise}$ is a measure for the noise contribution in the excited frequency bin, determined with the LPM for each excited frequency individually. Taking the average of the noise for all excited frequencies defines an average noise floor $\bar{\sigma}_{noise}$. Measurements with an SNR $\ge 5$ are considered to represent a reliable FRD point \cite{Berkel2015ThesisPhd}.

\section{Shot overview}

\subsection{D$_{2}$ fuelling perturbations}

\begin{table}[H]
    \centering
    \caption{Overview of discharges with D$_2$ gas fuelling perturbations. Excitation frequencies $f_{exc}$, amplitude $A$ of the designed signal, start time of the perturbations $t_{start}$, and amount of completed periods $P$ are given.}
    \label{tab: fuelling shotoverview}
    \begin{tabular}{|c|c|c|c|c|c|}
        \hline
        \textbf{Shot} & \textbf{Configuration} & \textbf{$f_{exc}$ [Hz]} & $A$ [$10^{20} \ \mathrm{part.}/\mathrm{s}$]&\textbf{$t_{start}$ [s]}& \textbf{$P$} \\ 
        \hline
        78131 & SN, Case A & $3.57 \cdot [1,3,7,9]$ & 4 & 0.9 & 3 \\
        \hline
        85292 & XPT, Case A & $6.25 \cdot [1]$ & 2 & 1 & 5 \\
        \hline
        85293 & XPT, Case A & $6.25 \cdot [1]$ & 2 & 1 & 5 \\
        \hline
        85295 & XPT, Case A & $6.25 \cdot [1,3,5]$ & 2 & 1 & 5 \\ 
        \hline
        85301 & XPT, Case A & $6.25 \cdot [1,3,5]$ & 2 & 1 & 4 \\ 
        \hline
        85340 & SN, Case A & $6.25 \cdot [1,3,5]$& 2 & 1 & 4 \\ 
        \hline
        85343 & XPT, Case A & $6.25 \cdot [1, 3]$ & 4 & 1 & 4 \\ 
        \hline
        85344 & SN, Case A & $6.25 \cdot [1,3,5]$ & 2 & 1 & 3 \\
        \hline
        85347 & SN, Case A & $4.35 \cdot [1,3]$ & 4 & 1 & 5 \\ 
        \hline
        85350 & XPT, Case A & $6.25 \cdot [1,3]$ & 4 & 1 & 4 \\
        \hline
        85353 & XPT, Case A & $4.35 \cdot [1,3]$ & 4 &1 & 4 \\ 
        \hline
    \end{tabular}
\end{table}


\subsection{N$_{2}$ seeding perturbations}

\begin{table}[h!]
    \centering
    \caption{Overview of discharges with N$_{2}$ gas seeding perturbations. Excitation frequencies $f_{exc}$, amplitude $A$ of the designed signal, start time of the perturbations $t_{start}$, and amount of completed periods $P$ are given.}
    \label{tab: seeding shotoverview}
    \begin{tabular}{|c|c|c|c|c|c|}
        \hline
        \textbf{Shot} & \textbf{Configuration} & \textbf{$f_{exc}$ [Hz]} & $A$ [$10^{20}$ part./s] &\textbf{$t_{start}$ [s]} & \textbf{$P$} \\ 
        \hline
        85380 & XPT, Case B & $6.25 \cdot [1,3]$ & 2.75 & 1.2 & 3  \\ 
        \hline
        85388 & SN, Case B &  $6.25 \cdot [1,3]$  & 2.75 & 1.2 & 3   \\ 
        \hline
    \end{tabular}
\end{table}

\subsection{ECRH power perturbations}

\begin{table*}[ht!]
    \centering
    \caption{Overview of discharges with ECRH power modulations. Excitation frequencies $f_{exc}$, amplitude $A$ of the designed signal, start time of the perturbations $t_{start}$, and amount of completed periods $P$ are given.}
    \label{tab: heating shotoverview}
    \begin{tabular}{|c|c|c|c|c|c|}
        \hline
        \textbf{Shot} & \textbf{Configuration} & \textbf{$f_{exc}$ [Hz]} & $A$ [kW] &\textbf{$t_{start}$ [s]} & \textbf{$P$} \\ 
        \hline
        85375 & XPT, Case B & [25.7, 44.6] & 300 & 1.2 & 3  \\ 
        \hline
        85377 & SN, Case B &  [25.7, 44.6] & 300 & 1.2 & 3   \\ 
        \hline
        89544 & SN, Case B &  [7.7, 38.5] & 300 & 1.2 & 3   \\ 
        \hline
    \end{tabular}
\end{table*}

\newpage

\begin{table}[ht]
\centering
\caption{SNR values of the FRD determined via the system identification method described in \ref{appendix: Frequency Response Data} in discharges with fuelling perturbations. A SNR$>5$ is considered to be a good measurement and is coloured green. Points with $3<$SNR$<5$ are coloured orange and are less accurate, yet still considered valuable. Points with SNR$<3$ are considered to be of poor quality. For entries with an 'x' the diagnostic was not available during that discharge. }
\label{tab: SNR table fuelling shots}
\small
\setlength{\tabcolsep}{5pt}
\renewcommand{\arraystretch}{1.15}
\begin{tabular}{|l|l|l|l|l|l|l|l|l|l|l|} 
\hline
\multicolumn{11}{|l|}{\textbf{SNR overview}} \\ \hline 
Shotnumbers & \#85292 & \#85293 & \#85295 & \#85301 & \#85340 & \#85343 & \#85344 & \#85347 & \#85350 & \#85353 \\ \hline
Configuration & XPT & XPT & XPT & XPT & SN & XPT & SN & SN & XPT & XPT \\ \hline
$f_0$         & 6.25 & 6.25 & 6.25 & 6.25 & 6.25 & 6.25 & 6.25 & 4.35 & 6.25 & 4.35 \\ \hline
$f_{\mathrm{exc}}$ & {[}1{]} & {[}1{]} & {[}1,3,5{]} & {[}1,3,5{]} & {[}1,3,5{]} & {[}1,3{]} & {[}1,3,5{]} & {[}1,3{]} & {[}1,3{]} & {[}1,3{]} \\ \hline

\multicolumn{11}{|l|}{\textbf{MANTIS: $L_{pol}$}} \\ \hline
{[}1{]} & 
{\color[HTML]{00B050} 5.3} & 
{\color[HTML]{FFC000} 3.3} &
{\color[HTML]{FF0000} 2.8} &
{\color[HTML]{00B050} 5.7} &
{\color[HTML]{FFC000} 4.8} &
{\color[HTML]{00B050} 5.8} &
{\color[HTML]{FF0000} 1.9} &
{\color[HTML]{00B050} 13.7} &
{\color[HTML]{00B050} 20.5} &
{\color[HTML]{00B050} 10.1} \\ \hline
{[}3{]} & & &
{\color[HTML]{00B050} 8.7} &
{\color[HTML]{00B050} 8.6} &
{\color[HTML]{00B050} 5.1} &
{\color[HTML]{00B050} 30.8} &
{\color[HTML]{00B050} 6.0} &
{\color[HTML]{00B050} 5.1} &
{\color[HTML]{00B050} 7.1} &
{\color[HTML]{00B050} 24.0} \\ \hline
{[}5{]} & & &
{\color[HTML]{FF0000} 1.1} &
{\color[HTML]{FF0000} 1.5} &
{\color[HTML]{FFC000} 3.1} &
 &
{\color[HTML]{00B050} 7.2} &
& & \\ \hline

\multicolumn{11}{|l|}{\textbf{PD: $D_{\alpha,\mathrm{top}}$}} \\ \hline
{[}1{]} & 
{\color[HTML]{00B050} 36.5}& 
{\color[HTML]{00B050} 19.7}&
{\color[HTML]{00B050} 13.4} &
{\color[HTML]{00B050} 20.7} &
{\color[HTML]{00B050} 12.2} &
{\color[HTML]{00B050} 29.4} &
{\color[HTML]{00B050} 8.0} & 
{\color[HTML]{00B050} 12.3} &
{\color[HTML]{00B050} 12.5} &
{\color[HTML]{00B050} 22.8} \\ \hline
{[}3{]} & 
& 
&
{\color[HTML]{00B050} 20.8} &
{\color[HTML]{00B050} 15.2} &
{\color[HTML]{00B050} 22.0} &
{\color[HTML]{00B050} 23.9} &
{\color[HTML]{00B050} 45.9} &
{\color[HTML]{00B050} 28.3} &
{\color[HTML]{00B050} 16.1} &
{\color[HTML]{00B050} 19.1} \\ \hline
{[}5{]} & & &
{\color[HTML]{00B050} 8.3} &
{\color[HTML]{00B050} 28.5} &
{\color[HTML]{00B050} 9.2} &
 &
{\color[HTML]{00B050} 15.0} &
& & \\ \hline

\multicolumn{11}{|l|}{\textbf{PD: $D_{\alpha,\mathrm{div}}$}} \\ \hline
{[}1{]} &
{\color[HTML]{00B050} 21.9}& 
{\color[HTML]{00B050} 43.8}&
{\color[HTML]{00B050} 50.1} &
{\color[HTML]{00B050} 192.9} &
{\color[HTML]{00B050} 18.8} &
{\color[HTML]{00B050} 40.6} &
{\color[HTML]{00B050} 13.4} &
{\color[HTML]{00B050} 28.6} &
{\color[HTML]{00B050} 61.0} &
{\color[HTML]{00B050} 171.1} \\ \hline
{[}3{]} & & &
{\color[HTML]{00B050} 62.2} &
{\color[HTML]{00B050} 28.6} &
{\color[HTML]{00B050} 35.9} &
{\color[HTML]{00B050} 162.8} &
{\color[HTML]{00B050} 41.4} &
{\color[HTML]{00B050} 66.4} &
{\color[HTML]{00B050} 50.8} &
{\color[HTML]{00B050} 41.4} \\ \hline
{[}5{]} & & &
{\color[HTML]{00B050} 52.3} &
{\color[HTML]{00B050} 51.2} &
{\color[HTML]{00B050} 22.3} &
 &
{\color[HTML]{00B050} 127.6} &
& & \\ \hline

\multicolumn{11}{|l|}{\textbf{APG: $p_{0,\mathrm{div}}$}} \\ \hline
{[}1{]} & 
x& 
x&
x & x &
{\color[HTML]{FF0000} 2.4} &
{\color[HTML]{00B050} 21.5} &
{\color[HTML]{FFC000} 3.4}&
{\color[HTML]{FFC000} 3.7} &
{\color[HTML]{00B050} 7.7} &
{\color[HTML]{00B050} 5.9} \\ \hline
{[}3{]} & & &
x & x &
{\color[HTML]{FFC000} 3.8} &
{\color[HTML]{00B050} 10.6} &
{\color[HTML]{FFC000} 4.5} &
{\color[HTML]{00B050} 11.5} &
{\color[HTML]{00B050} 7.9} &
{\color[HTML]{00B050} 12.9} \\ \hline
{[}5{]} & & &
x & x &
{\color[HTML]{00B050} 5.9} &
 &
{\color[HTML]{FF0000} 2.0} &
& & \\ \hline

\multicolumn{11}{|l|}{\textbf{APG: $p_{0,\mathrm{bot}}$}} \\ \hline
{[}1{]} & x & x &
x & x &
{\color[HTML]{00B050} 6.0} &
{\color[HTML]{00B050} 7.8} &
{\color[HTML]{00B050} 6.1} &
{\color[HTML]{FFC000} 3.0} &
{\color[HTML]{00B050} 6.0} &
{\color[HTML]{00B050} 6.7} \\ \hline
{[}3{]} & & &
x & x &
{\color[HTML]{FFC000} 4.7} &
{\color[HTML]{00B050} 7.2} &
{\color[HTML]{FFC000} 3.6} &
{\color[HTML]{00B050} 5.2} &
{\color[HTML]{00B050} 5.0} &
{\color[HTML]{00B050} 5.5} \\ \hline
{[}5{]} & & &
x & x &
{\color[HTML]{FF0000} 1.0} &
 & 
{\color[HTML]{FF0000} 1.9} &
& & \\ \hline

\multicolumn{11}{|l|}{\textbf{FIR: $n_{\text{core}}$}} \\ \hline
{[}1{]} & 
{\color[HTML]{00B050} 7.5}& 
{\color[HTML]{00B050} 10.7}&
{\color[HTML]{00B050} 6.3} &
{\color[HTML]{00B050} 26.1} &
{\color[HTML]{00B050} 9.4} &
{\color[HTML]{00B050} 25.1} &
{\color[HTML]{FF0000} 2.4} &
{\color[HTML]{00B050} 5.7} &
{\color[HTML]{00B050} 358.9} &
{\color[HTML]{FF0000} 2.6} \\ \hline
{[}3{]} & & &
{\color[HTML]{00B050} 8.2} &
{\color[HTML]{00B050} 9.8} &
{\color[HTML]{00B050} 6.9} &
{\color[HTML]{00B050} 19.3} &
{\color[HTML]{FFC000} 3.7} &
{\color[HTML]{00B050} 21.8} &
{\color[HTML]{00B050} 9.2} &
{\color[HTML]{00B050} 17.9} \\ \hline
{[}5{]} & & &
{\color[HTML]{FFC000} 4.2} &
{\color[HTML]{FF0000} 1.9} &
{\color[HTML]{FF0000} 1.2} &
 &
{\color[HTML]{FF0000} 1.1}&
& & \\ \hline
\end{tabular}
\end{table}

\newpage
\section{Extended FRD analysis}\label{appendix: Sect, Extended FRD analysis}
\subsection{XPT FRD analysis per region}\label{appendix: Small amplitude perturbations}

\begin{figure}[h]
    \centering
    \includegraphics[width=\linewidth]{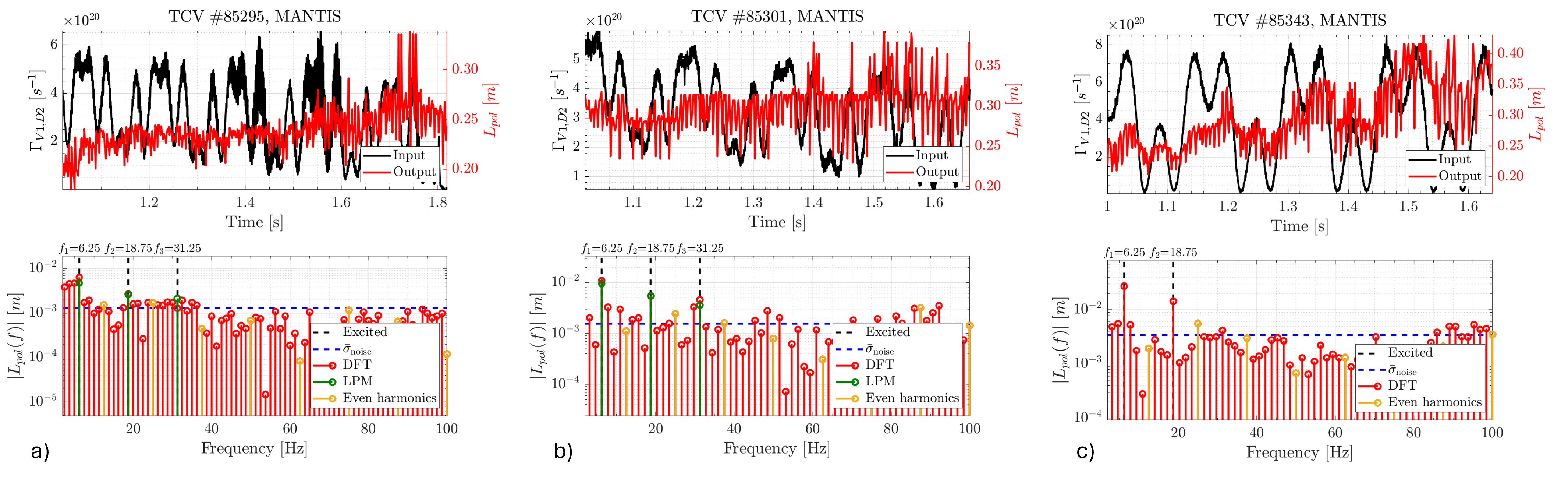}
    \caption{Response in time (top) and frequency (bottom) domain of the CIII front to D$_2$ fuelling perturbations in discharges XPT 85295, 85301 and 85343. The excited frequencies (black, dashed), average noise floor (blue, dashed), DFT (red), LPM correction of the DFT (green) and even multiples of the excited ground frequency $f_0 = 6.25$ Hz (orange) are shown. No dominant even non-linearities are detected in this frequency range for all discharges, indicating a dominantly linear response. A minimal response to the excited frequencies is observed in (a), where the operating point of the CIII front is close to the secondary X-point, whereas the exited frequencies in (b) and (c) do show a response.}
    \label{fig: Small amplitude pert Fueling}
\end{figure}


\newpage
\newpage
\subsection{Multi-output analysis}\label{appendix: multi output}
\begin{figure}[b] 
    \centering
    \includegraphics[width=0.8\linewidth]{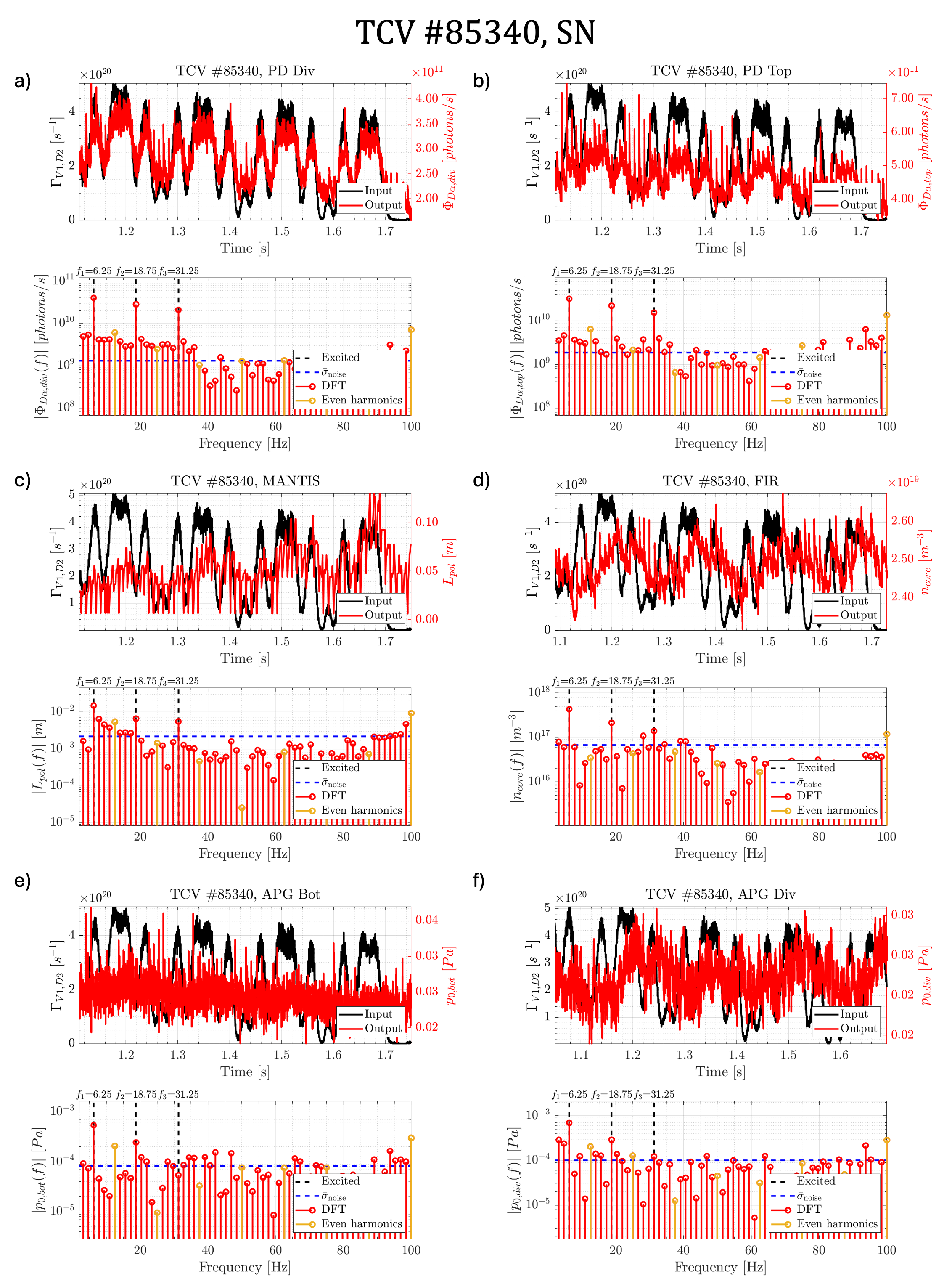}
    \caption{Response in time and frequency domain of various diagnostics to D$_2$ gas perturbations via Valve 1 in TCV discharge 85340, with a SN divertor. a) and b) show the response of the $D_{\alpha}$ by photodiodes with respectively divertor, and top-bottom viewlines. c) Shows the response of the CIII front $L_{pol}$ measured by MANTIS, d) the response of the core density $n_{core}$ measured by FIR, e) and f) the neutral pressures $p_{0,bot}$ and $p_{0,div}$ of APG at 2 locations.  }
    \label{fig: Linearity 85340}
\end{figure}
\clearpage 

\begin{figure} 
    \centering
    \includegraphics[width=0.8\linewidth]{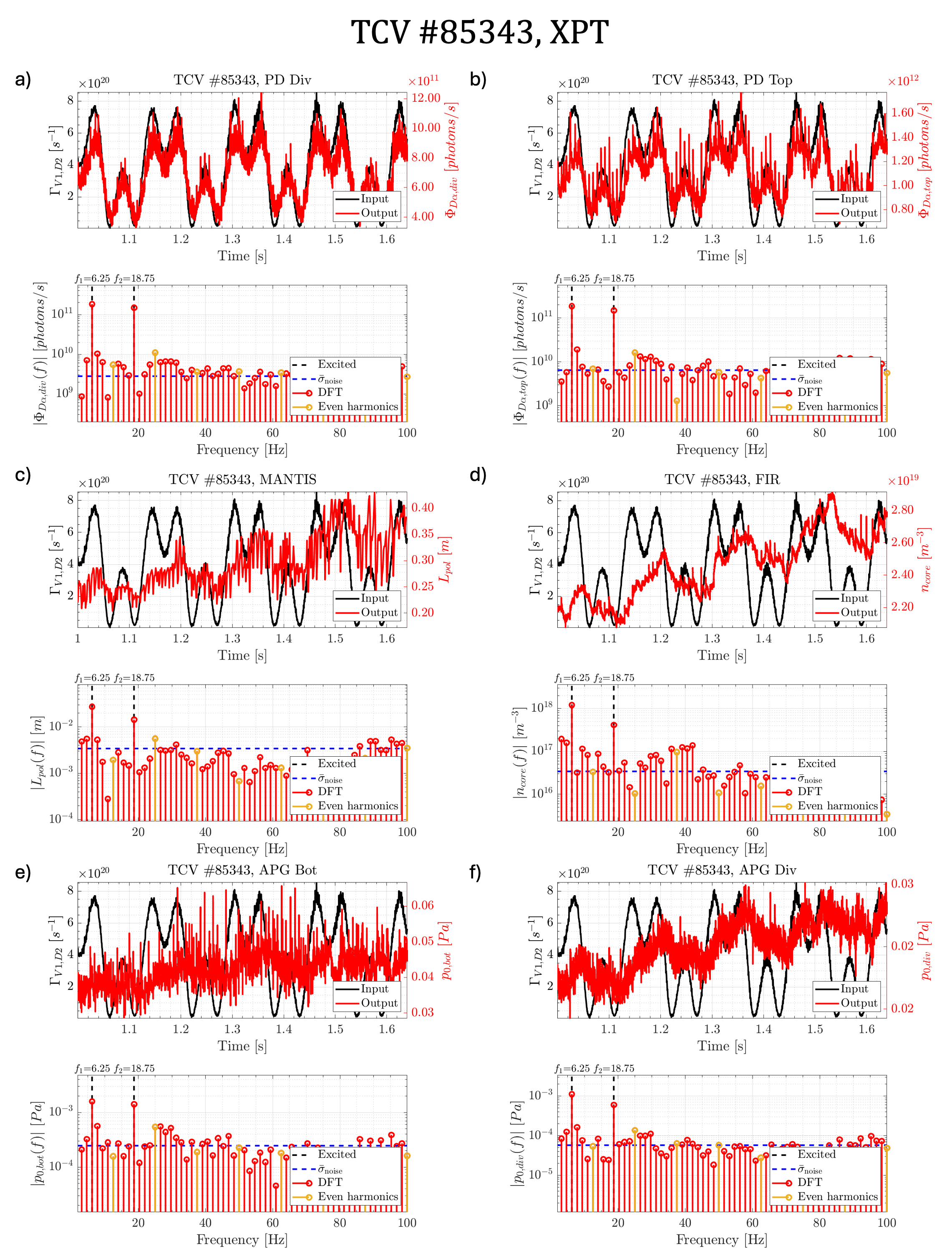}
    \caption{Response in time and frequency domain of various diagnostics to D$_2$ gas perturbations via Valve 1 in TCV discharge 85343, with a XPT divertor. a) and b) show the response of the $D_{\alpha}$ by photodiodes with respectively divertor, and top-bottom viewlines. c) Shows the response of the CIII front $L_{pol}$ measured by MANTIS, d) the response of the core density $n_{core}$ measured by FIR, e) and f) the neutral pressures $p_{0,bot}$ and $p_{0,div}$ of APG at 2 locations.  }
    \label{fig: Linearity 85343}
\end{figure}
\clearpage 


\begin{figure} 
    \centering
    \includegraphics[width=0.8\linewidth]{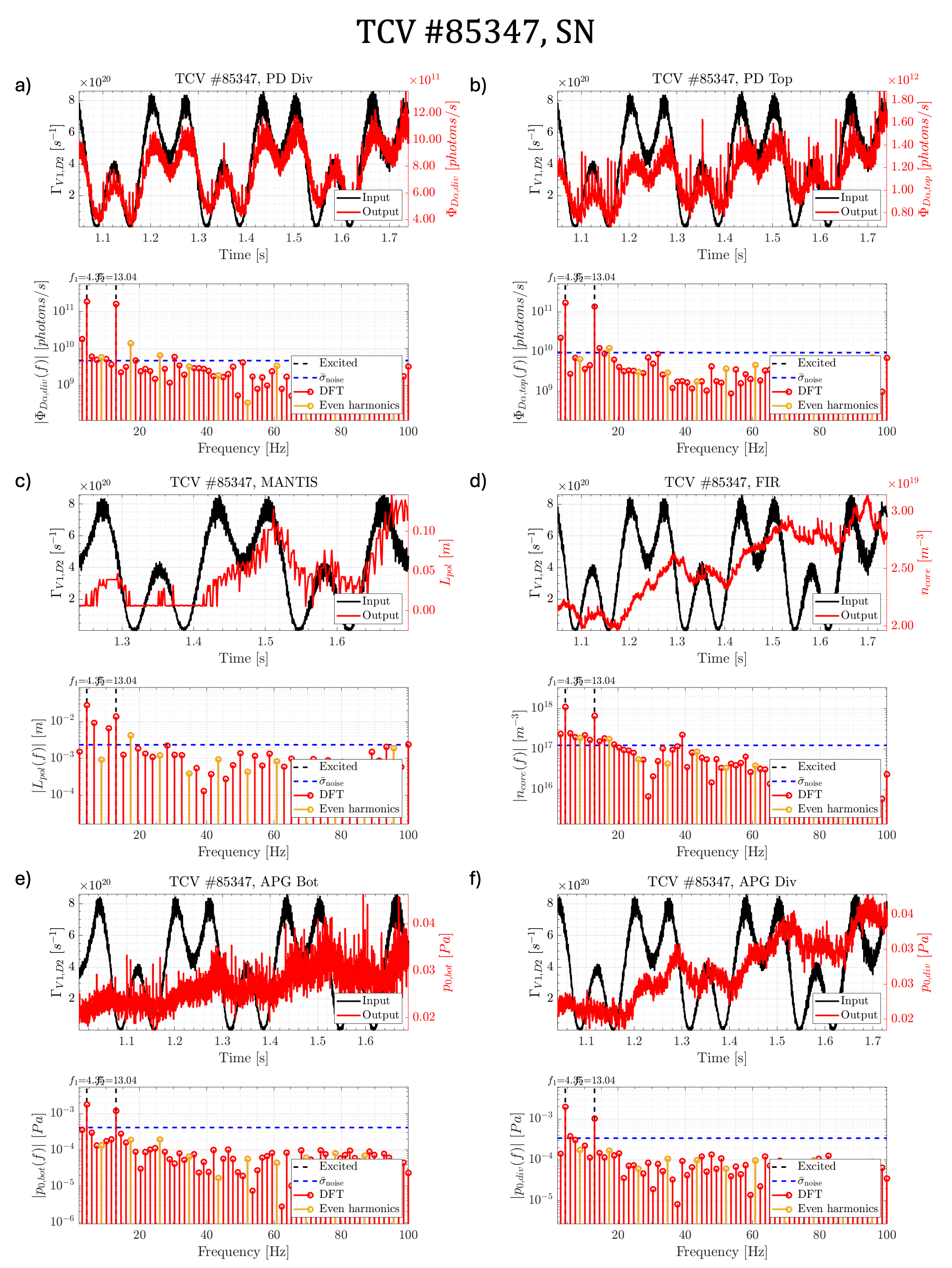}
    \caption{Response in time and frequency domain of various diagnostics to D$_2$ gas perturbations via Valve 1 in TCV discharge 85343, with a SN divertor. a) and b) show the response of the $D_{\alpha}$ by photodiodes with respectively divertor, and top-bottom viewlines. c) Shows the response of the CIII front $L_{pol}$ measured by MANTIS, d) the response of the core density $n_{core}$ measured by FIR, e) and f) the neutral pressures $p_{0,bot}$ and $p_{0,div}$ of APG at 2 locations.  }
    \label{fig: Linearity 85347}
\end{figure}
\clearpage 

\begin{figure}
    \centering
    \includegraphics[width=0.8\linewidth]{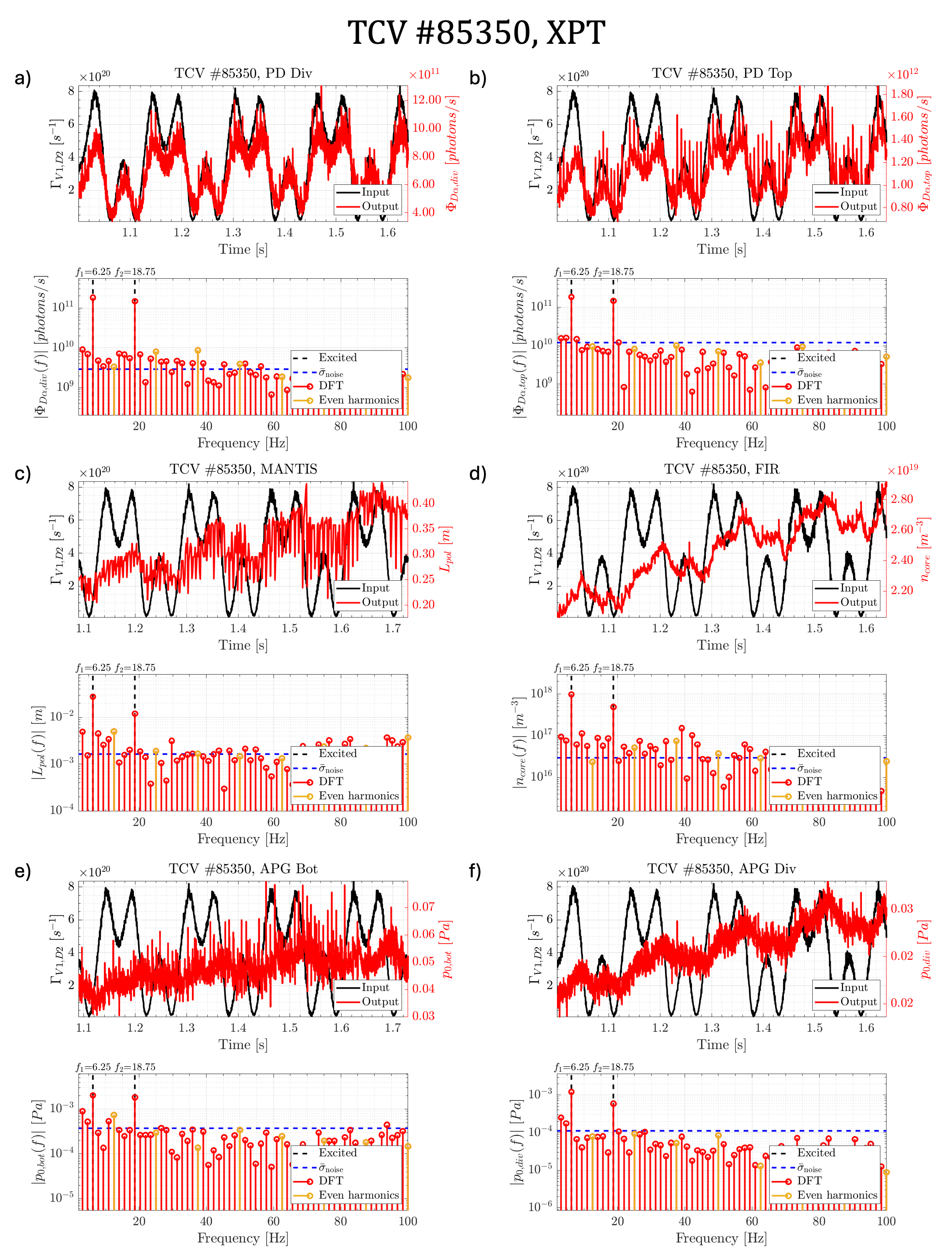}
    \caption{Response in time and frequency domain of various diagnostics to D$_2$ gas perturbations via Valve 1 in TCV discharge 85350, with a XPT divertor. a) and b) show the response of the $D_{\alpha}$ by photodiodes with respectively divertor, and top-bottom viewlines. c) Shows the response of the CIII front $L_{pol}$ measured by MANTIS, d) the response of the core density $n_{core}$ measured by FIR, e) and f) the neutral pressures $p_{0,bot}$ and $p_{0,div}$ of APG at 2 locations.  }
    \label{fig: Linearity 85350}
\end{figure}
\clearpage

\begin{figure} 
    \centering
    \includegraphics[width=0.8\linewidth]{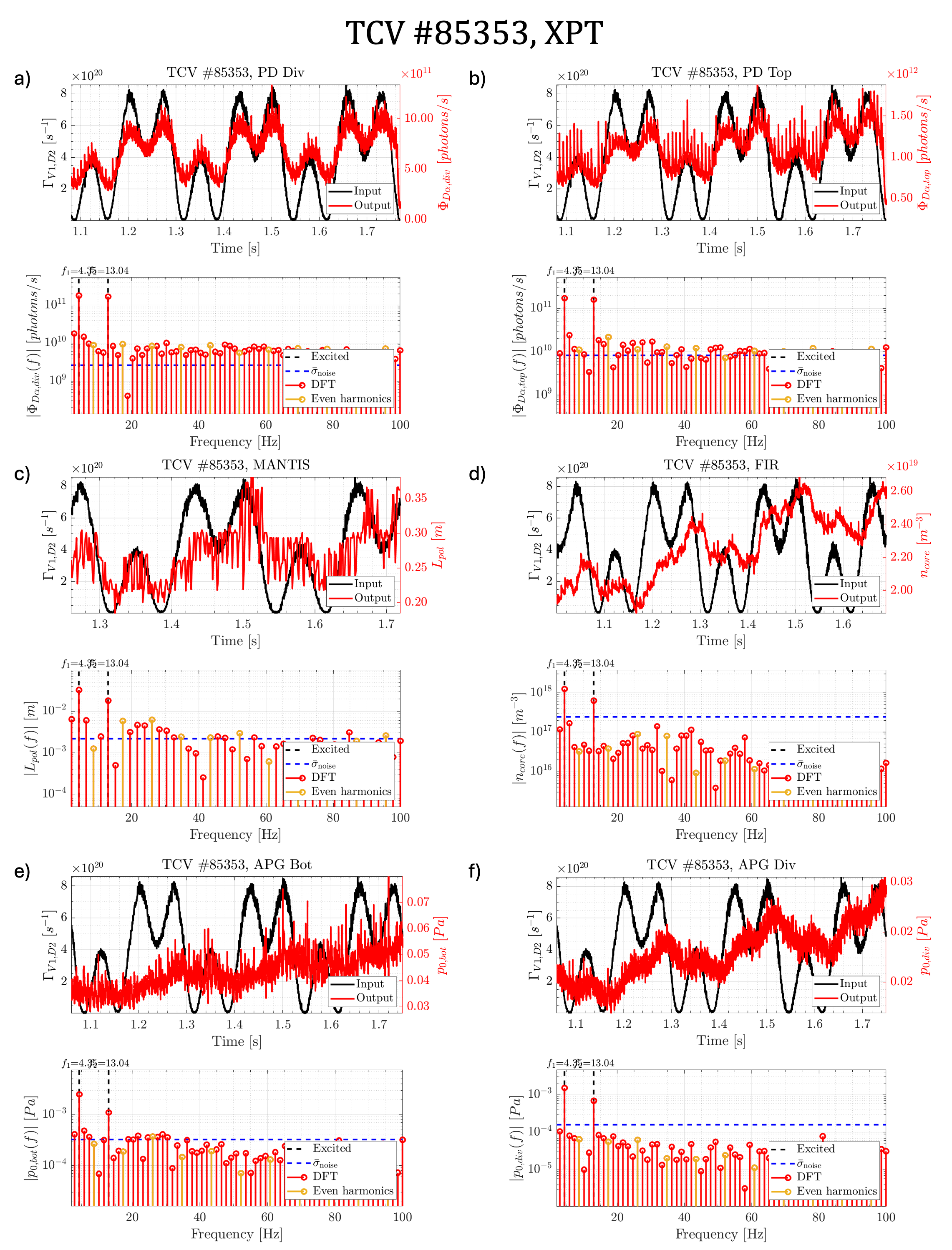}
    \caption{Response in time and frequency domain of various diagnostics to D$_2$ gas perturbations via Valve 1 in TCV discharge 85353, with a XPT divertor. a) and b) show the response of the $D_{\alpha}$ by photodiodes with respectively divertor, and top-bottom viewlines. c) Shows the response of the CIII front $L_{pol}$ measured by MANTIS, d) the response of the core density $n_{core}$ measured by FIR, e) and f) the neutral pressures $p_{0,bot}$ and $p_{0,div}$ of APG at 2 locations.  }
    \label{fig: Linearity 85353}
\end{figure}



\clearpage

\section{DLS}\label{appendix: DLS}
The Detachment Location Sensitivity (DLS) model \cite{Lipschultz2016_DLS1,Myatra2023DLS} defines an infinitely thin impurity radiation front along a flux tube in the SOL that spans the outboard midplane to the outer divertor target. Along the fluxtube, a parallel coordinate $z$ is defined as 

\begin{equation}\label{eq: DLS coordinate}
    dz = \frac{B_{\times}}{B(L)} dL = \frac{B_{x}}{B_{pol}(L)}dL_{pol},
\end{equation}

Here $B_{\times}$ is the magnetic field at the primary X-point, $B_{pol}$ the poloidal component of the magnetic field along the fluxtube, $L$ the non-scaled parallel coordinate along the flux tube, and $L_{pol}$ the poloidal projection of the parallel coordinate. \par
Within the fluxtube that represents the entire divertor leg, the impurity front is assumed to be responsible for all power dissipation via impurity radiation with a single impurity specie. Hence, momentum, particle and power losses via ionization and plasma-neutral interactions are ignored. According to the DLS model, the location of the front is determined by three factors.\par
\begin{enumerate}
\item{A set of three plasma parameters. The plasma parameters are lumped together in a driver parameter $D = \frac{n_{u} f_{z}^{1/2}}{q_{\parallel,u}^{5/7}}$, with $n_{u}$ the upstream electron density at the outboard midplane, $f_{z}$ the impurity fraction, and $q_{\parallel,u}$ the upstream parallel heat flux.}
\item{The magnetic geometry upstream of the front. A single magnetic field line from the outboard midplane to the front location along the divertor leg is chosen. This is a limitation for configurations with multiple strike points such as the XPT. There, heat and particle fluxes are guided to multiple strike points in different spatial regions (e.g. SP2 and SP4 in Figure \ref{fig: experimental setup}), with varying magnetic geometries. Therefore, this work only considers the magnetic field line that carries the dominant heat flux contribution for the chosen equilibria, e.g. a field line towards SP4 in the presented XPT discharges.}
\item{A 'cooling function' which is assumed to be constant and invariant of the chosen magnetic geometry. Therefore, the analysis in this Section only considers the first two points as a function of the front position $L_{pol}$.}
\end{enumerate}
This is combined in a single equation given by: 

\begin{equation}
    \frac{n_u \sqrt{f_{z}}}{q_{\parallel,u}^{5/7}} = \frac{1}{U}\frac{B(z_f)}{B_{\times}^{3/7}}\times \left[ \int_{z_f}^{z_{\times}}B^2(z) dz + \int_{z_{\times}}^L \frac{B^2(z)(L-z)}{L-z_{\times}}dz\right]^{-2/7}
\end{equation}

Here $U$ is assumed to be a constant cooling function that originates from the Lengyel integral \cite{Lengyel1981LengyelIntegral}, given by

\begin{equation}
    U = 7^{2/7}(2\kappa_0)^{3/14}\sqrt{\int T^{1/2}Q(T)dT},
\end{equation}

with a cooling curve of the radiating specie $Q(T)$, $T$ the temperature, and $\kappa_0$ the heat conduction coefficient of the plasma.

\newpage

\bibliography{referencesXPT}

\end{document}